\documentclass[10pt]{amsart}

\usepackage[dvips]{graphicx}
\usepackage{psfrag}
\usepackage{amsmath,amsthm,amssymb}
\usepackage{latexsym}
\usepackage{enumerate}
\newcounter{satz}

\newtheorem{theorem}[satz]{Theorem}
\newtheorem{lemma}{Lemma}[section]
\newtheorem{prop}[lemma]{Proposition}

\newenvironment{specprop}[1]%
{\smallskip \begin{sloppypar}\noindent\bf Proposition #1. \it}%
{\end{sloppypar}}

\newtheorem{cor}[lemma]{Corollary}
\newtheorem{dfn}[lemma]{Definition}

\newcommand{\vol}{{\rm vol}}
\newcommand{\inj}{{\rm inj}}

\newcommand{\RR}{{\mathbb R}}
\newcommand{\HH}{{\mathbb H}}

\begin{document}

\title{Geometric heat comparison criteria for Riemannian manifolds}

\author{Leon Karp \and Norbert Peyerimhoff}
\date{May 26, 2005}

%%%%%%%%%%%%%%%%%%%%%%%%%%%%%%%%%%%%%%%%%%%%%%%%%%%%%%%%%%%%%%%%%%%%%%%%%%%

\begin{abstract} \noindent
The main results of this article are {\em small time} heat comparison
results for two points in two manifolds with characteristic functions
as initial temperature distributions (Theorems 1 and 2). These results
are based on the geometric concepts of {\em (essential) distance from the
complement} and {\em spherical area function}. We also discuss some
other geometric results about the heat development and illustrate them
by examples.
\end{abstract}

\maketitle

\renewcommand{\thefootnote}{\fnsymbol{footnote}}
\footnotetext[0]{{\em Mathematics Subject Classification} (2000): 58J35, 35K05}

%%%%%%%%%%%%%%%%%%%%%%%%%%%%%%%%%%%%%%%%%%%%%%%%%%%%%%%%%%%%%%%%%%%%%%%%%%%

\section{Introduction: Examples and statement of results}

This article is mainly concerned with small time properties of the
heat flow on Riemannian manifolds. Of particular interest are {\em
geometric} heat comparison criteria for two different points in two
manifolds. All Riemannian manifolds $M$ considered in this Introduction are
connected, complete and without boundary. We also assume that they
have a lower (not necessarily positive) bound on the Ricci curvature,
an upper bound on the sectional curvature and a positive lower bound
on the injectivity radius. For any closed subset $\Omega \subset M$ with 
$\vol_n(\partial \Omega) = 0$ (where $n = \dim M$ and $\vol_n$ 
denotes the Riemannian measure on $M$) let $f_{\Omega,X}: (0,\infty) 
\times M \to \RR$ denote
the smooth solution of the heat equation
\begin{equation} \label{heateq}
\frac{\partial}{\partial t} f(t,x) = \Delta f(t,x), \quad \lim_{t \to 0^+}
f(t,x) = \chi_\Omega(x) \quad \text{for all $x \in M \backslash \partial 
\Omega$},
\end{equation}
where $\chi_\Omega$ is the characteristic function of $\Omega$.
The solution $f_{\Omega,M}$ is given by
\[ f_{\Omega,M}(t,x) = \int_\Omega k_M(t,x,y) dy, \]
where $k_M$ is the heat kernel on $M$. $f_{\Omega,M}(t,\cdot)$
describes the temperature distribution of the heat flow at time $t >
0$ for the given initial temperature distribution $\chi_\Omega$. Our
comparison data are given by triples $(x,\Omega,M)$, where $M$ is a
Riemannian manifold (with the above properties), $x \in M$ and
$\Omega$ is a closed subset of $M$ with $\vol_n(\partial \Omega)=0$. 
A closed subset $\Omega \subset M$ with $\vol_n(\partial \Omega) = 0$ is 
called henceforth {\em admissible}.

\begin{dfn}
We say that $(x_1,\Omega_1,M_1)$ is {\em initially hotter} than
$(x_2,\Omega_2,M_2)$ if there exists a time $\tau > 0$ such that
\begin{equation} \label{inhot}
f_{\Omega_1,M_1}(t,x_1) \ge f_{\Omega_2,M_2}(t,x_2) \quad
\text{for all $t \in (0,\tau)$}.
\end{equation}
$(x_1,\Omega_1,M_1)$ is {\em initially strictly hotter} than
$(x_2,\Omega_2,M_2)$, if inequality \eqref{inhot} holds
strictly.

If $(x_1,\Omega_1,M_1)$ is initially hotter, resp., initially
strictly hotter than $(x_2,\Omega_2,M_2)$, we write shortly
\[
(x_1,\Omega_1,M_1) \succeq (x_2,\Omega_2,M_2), \
\text{respectively},\ (x_1,\Omega_1,M_1) \succ
(x_2,\Omega_2,M_1).
\]
\end{dfn}

We also want to compare the initial temperatures of {\em subsets} of two 
manifolds.

\begin{dfn}
We say that the set $I_1 \subset M_1$ is {\em uniformly
initially strictly hotter} than $I_2 \subset M_2$, if for all $x_1
\in I_1$ and $x_2 \in I_2$ inequality \eqref{inhot} holds
strictly with a uniform $\tau > 0$. In this case we write 
$(I_1,\Omega_1,M_1) \succ (I_2,\Omega_2,M_1)$.
\end{dfn}

To state our first result, we need some preparations. Let $\Omega
\subset M$ be an admissible subset. Let $B_r(x)$ denote
the closed ball of radius $r$ about $x$. The function
$R: M \to [0,\infty]$, defined by
\[
R(x) := \sup\{ r \ge 0 \mid \vol_n(B_r(x)) = \vol_n(B_r(x)\cap \Omega) \},
\]
is called the {\em (essential) distance of $x$ from the
complement $\Omega^c = M \backslash \Omega$}. Note that $x \in 
\overline{\Omega^c}$ implies $R(x) = 0$. Moreover, $R(y) \ge R(x) - d(x,y)$ 
implies that $R$ is continuos. If $\Omega$ is a closed set with piecewise 
smooth boundary then we have
\[ R(x) = d(x,\overline{\Omega^c}). \]
The supremum
\[ R_\infty(\Omega) := \sup_{x \in \Omega} R(x) \]
is called the {\em inradius of the set $\Omega$} and the set
\[ I_\infty(\Omega) := \{ x \in \Omega \mid R(x) = R_\infty(\Omega)
\}\] 
denotes the set of {\em maximally interior points} of $\Omega$.
Continuity of $R: M \to [0,\infty]$ implies that
$I_\infty(\Omega)$ is a closed set.

\medskip

Let $\inj(x)$ denote the injectivity radius of $M$ at $x$. The
distance from the complement plays an important role in the following
comparison criterion.

\begin{theorem} \label{theo1}
Let $(x_1,\Omega_1,M_1)$ and $(x_2,\Omega_2,M_2)$ be given and let 
$R_j: M_j \to [0,\infty]$ be the corresponding distances from the 
complements. If $R_j(x_j) < \inj(x_j)$, for $j = 1,2$, and
\[ R_1(x_1) < R_2(x_2), \]
then $(x_2,\Omega_2,M_2)$ is initially strictly hotter than
$(x_1,\Omega_1,M_1)$.

More general, given $0 \le R_1 < R_2$ and two compact sets $I_1 \subset M_1$
and $I_2 \subset M_2$ satisfying
\begin{itemize}
\item[(a)] $R_1(x_1) \le R_1$ for all $x_1 \in I_1$
and $R_2(x_2) \ge R_2$ for all $x_2 \in I_2$,
\item[(b)] $R_2 < \inj(x)$ for all $x \in I_1 \cup I_2$.
\end{itemize}
Then $I_1$ is uniformly initially strictly hotter than $I_2$. 
\end{theorem}

Note that if $\Omega \subset M$ is an admissible set then so is
$\overline{\Omega^c}$. Using this fact and heat conservation (see property
(HK2) in Section 2), Theorem \ref{theo1} can also be used to compare points 
outside the domains $\Omega_j$.

\begin{cor} \label{theo1comp}
Let $(x_1,\Omega_1,M_1)$ and $(x_2,\Omega_2,M_2)$ be given and 
\[ R_j^-(x) := \sup\{r \ge 0 \mid \vol_n(B_r(x) \cap \Omega_j) = 0 \} \quad
\text{for $x \in M_j$}. \]
If $R_j^-(x_j) < \inj(x_j)$, for $j = 1,2$, and
\[ R_1^-(x_1) > R_2^-(x_2), \]
then $(x_2,\Omega_2,M_2)$ is initially strictly hotter than
$(x_1,\Omega_1,M_1)$.
\end{cor} 

The following example is an easy application of the theorem.

\medskip

\noindent {\bf Example 1:} Let $\Pi_1 \subset \RR^2$ and $\Pi_2
\subset \HH^2$ be two regular $n$-gons incribed in a Euclidean and a
hyperbolic circle of the same radius $R > 0$. Let $x_1 \in \Pi_1$ and $x_2
\in \Pi_2$ be the corresponding centers. Then the radii $r_1$ and $r_2$
of the corresponding inballs are given by
\[ r_1 = \cos(\pi/n) R \quad \text{and}\ \tanh r_2 = \cos(\pi/n) \tanh R, \]
and strict concavity of $r \mapsto \tanh(r) $ on $[0,\infty)$ implies
that $r_1 > r_2$. Hence, $(x_1,\Pi_1,\RR^2)$ is initially strictly
hotter than $(x_2,\Pi_2,\HH^2)$.

\medskip

Another consequence of Theorem \ref{theo1} is the existence of a
unique initially hottest point if there is a unique point $x \in
\Omega$ with largest distance to the boundary (see Corollary
\ref{uniqhot} below). Initially hottest points are defined as follows:

\begin{dfn}
Let $\Omega \subset M$ be an admissible subset. $x \in M$ is an 
{\em initially hottest point of $\Omega$} if and only if
\[
(x,\Omega,M) \succeq (x',\Omega,M) \quad \text{for all $x' \in
M$}.
\]
\end{dfn}

\begin{cor} \label{uniqhot}
Let $\Omega \subset M$ be an admissible subset and $R: M \to [0,\infty]$ 
be the corresponding distance from the complement. We
assume that $R(x) < \inj(x)$ for all $x \in \Omega$. If the set
$I_\infty(\Omega)$ of maximally interior points consists of only one
point then this point is also a unique initially hottest point of
$\Omega$.
\end{cor}

Corollary \ref{uniqhot} applies, e.g., to strictly convex compact
subsets $\Omega$ of $\RR^n$.

\smallskip

In \cite{ChK-90}, Chavel and Karp study the behaviour of the set
of hottest points
\[
H(t) = H_\Omega(t) =: \{ x_0 \in M \mid f_{\Omega,M}(t,x_0) =
\max_{x \in M} f_{\Omega,M}(t,x) \},
\]
as $t \to \infty$. The following result gives informations about
the set $H(t)$, as $t \to 0$.

\begin{cor} \label{conv0}
Let $\Omega \subset M$ be a compact admissible set of positive
volume, let $R: M \to [0,\infty)$ be the corresponding distance from the 
complement, and $H(t)$ be the set of hottest
points. We assume that $R_\infty(\Omega) < \inj(x)$ for all $x \in
\Omega$. Then we have, for any sequence $\{ x_j \}$ with $x_j \in
H(t_j)$, and $t_j \to 0$:
\[ d(x_j,I_\infty(\Omega)) \to 0. \]
as $j \to \infty$.
\end{cor}

Corollaries \ref{uniqhot} and \ref{conv0} are used in our next
example.

\medskip

\noindent {\bf Example 2:} 
Let $\Delta \subset \RR^2$ be an arbitrary Euclidean triangle. The
center of the inball of $\Delta$ is the unique initially hottest
point, by Corollary \ref{uniqhot}.  The results in \cite{ChK-90} imply
that the set of hottest points $H(t)$ remains in the triangle $\Delta$
for all $t > 0$; moreover, $H(t)$ converges to the center of mass of
$\Delta$, as $t \to \infty$. In combination with Corollary
\ref{conv0}, we conclude that the map $t \mapsto H(t)$ evolves from
the center of the inball and, finally, collapses into the center of
mass of the triangle. The precise trajectory of this map is not clear
to us. Numerical analysis shows that
the trajectory stays close to (but not on) the straight
Euclidean arc connecting these two centers. The experiments
seem also to indicate that $H(t)$ is always a single point, but we
lack a proof of this assumption. However, the next proposition implies
that $H(t)$ is a single point at least for sufficiently large $t$.

\begin{prop} \label{uniqpt}
Let $\Omega \subset \RR^n$ be a compact admissibe set of positive volume. 
Then there is a $T = T(\Omega) > 0$ such that $H(t)$
consists of a single point, for all $t \ge T$. Moreover, the map $t
\mapsto H(t)$ is a smooth curve and the temperature $t \mapsto
f_{\Omega,\RR^n}(t,H(t))$ at the hottest point is strictly decreasing
on the interval $(T,\infty)$.
\end{prop}

The monotonicity statement of Proposition \ref{uniqpt} and the proof
of it has been pointed out to us by Oliver Stein. Further applications
of Proposition \ref{uniqpt} are discussed in the next two examples.

\begin{figure}
\begin{center}
\psfrag{x0}{$z_0$} \psfrag{x1}{$z_1$} \psfrag{x2}{$z_2$}
\psfrag{x3}{$z_3$} \psfrag{L}{\LARGE $\Lambda$} \psfrag{D}{\LARGE
$\Delta$}
\includegraphics[width=\textwidth]{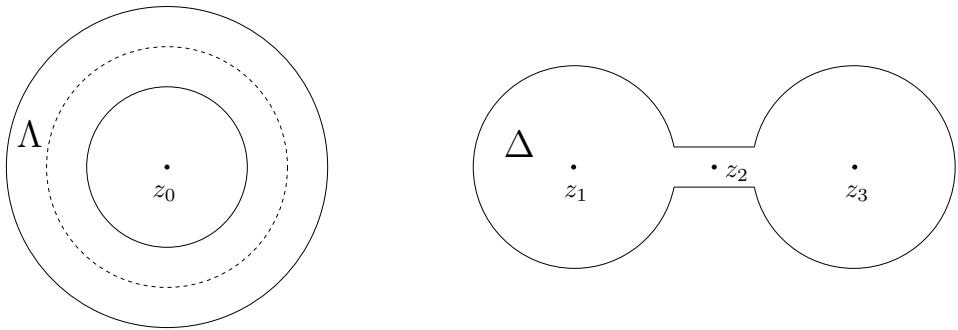}
\end{center}
\caption{Examples 3 and 4: Annulus and dumbbell in $\RR^2$}
\label{AnDb}
\end{figure}

\medskip

\noindent {\bf Examples 3 and 4:} Let $\Lambda$ and $\Delta$ be an
Euclidean annulus and a dumbbell, as presented in Figure
\ref{AnDb}.

The initially hottest points of the annulus $\Lambda$ form the
dashed circle. By symmetry reasons, the set $H(t)$ of hottest
points is spherically symmetric with respect to $z_0$, for all $t
> 0$. By \cite{ChK-90}, $H(t)$ shrinks to
$z_0$, as $t \to \infty$. Now, Proposition \ref{uniqpt} implies
that $H(t)$ arrives at $z_0$ {\em in finite time}.

$z_1$ and $z_3$ are the initially hottest points of the dumbbell
$\Delta$. Assume that the coordinates of $z_1$ and $z_3$ are
$(-a,0)$ and $(a,0)$. An easy argument, using the form of the heat
kernel along the vertical axes shows for any point $z = (x,y)$, $y
\neq 0$ that the point $(x,0)$ is initally strictly hotter than
$z$. This implies that $H(t) \subset \RR \times \{ 0 \}$, for all
times $t > 0$. By the same reasoning as above we conclude that
there is a finite time $T > 0$ such that $H(t) = \{ z_2 \}$ for
all $t > T$. The development of hottest points of a more simple dumbbell 
(with square ends) is explicitely discussed in Appendix \ref{sqdumbbell}.

\medskip

Note that there is no analogue of Proposition \ref{uniqpt} in
hyperbolic space. It was pointed out in \cite{ChK-90} that, for
$\Omega$ equals a large enough hyperbolic dumbbell, the set $H(t)$
does not converge to a single limit point, as $t \to \infty$.

\medskip

Our next result refers to Euclidean and hyperbolic polygons.

\begin{prop} \label{applmom}
Let $M$ be the Euclidean or the hyperbolic plane and $\Sigma_1,
\Sigma_2 \subset M$ be two polygons of the same area and the same
number of sides. Assume that $\Sigma_1$ is regular and that $x_1$
is its center. Then we have
\begin{equation} \label{polineq}
f_{\Sigma_1,M}(t,x_1) \ge \max_{x \in M} f_{\Sigma_2,M}(t,x).
\end{equation}
If \eqref{polineq} holds with equality for some time $t > 0$ then
$\Sigma_2$ is also regular. In particular, we have $(x_1,\Sigma_1,M)
\succeq (x,\Sigma_2,M)$, for all $x \in M$.
\end{prop}

Now, we move on to examples in which Theorem \ref{theo1} is not
applicable.

\medskip

\begin{figure}[h]
\begin{center}
\psfrag{x1}{$z_1$} \psfrag{x2}{$z_2$} \psfrag{x3}{$z_3$}
\psfrag{r0}{$R$} \psfrag{d}{$\delta$} \psfrag{O1}{\LARGE
$\Omega_1$} \psfrag{O2}{\LARGE $\Omega_2$}
\psfrag{>d}{$> \delta$}
\includegraphics[width=\textwidth]{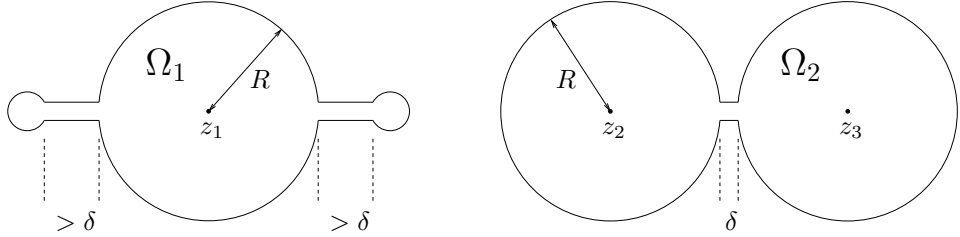}
\end{center}
\caption{Example 5: Comparison of $(z_1,\Omega_1,\RR^2)$ and
$(z_2,\Omega_2,\RR^2)$ } \label{ex5}
\end{figure}

\noindent {\bf Example 5:} In Figure \ref{ex5} the inradius at the
point $z_1$ in $\Omega_1$ coincides with the inradii at the points
$z_2$ and $z_3$ in $\Omega_2$. Therefore, the initial heat of
$z_1$ and $z_2$ cannot be compared with the help of Theorem
\ref{theo1}.

\medskip

\begin{figure}
\begin{center}
\psfrag{2R}{$2R$} 
\psfrag{z1}{$z_1$} 
\psfrag{z2}{$z_2$}
\psfrag{C}{$C$} 
\psfrag{S}{$\Sigma$} 
\includegraphics[width=\textwidth]{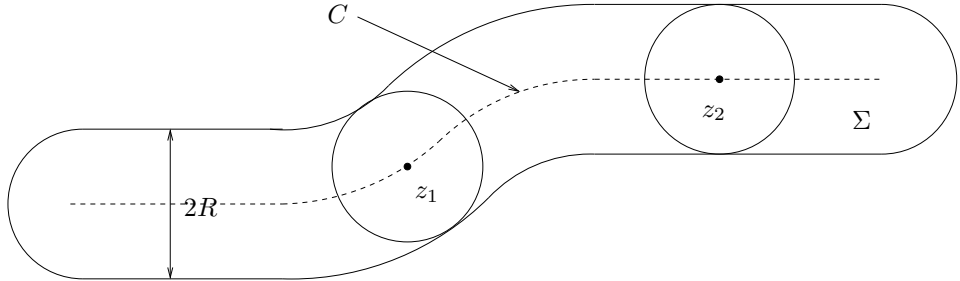}
\end{center}
\caption{Example 6: Comparison of $(z_1,\Sigma,\RR^2)$ and
$(z_2,\Sigma,\RR^2)$ } \label{ex6}
\end{figure}

\noindent {\bf Example 6:} Let $C \subset \RR^2$ be a plane
curve with absolute curvature bounded from above by a positive
constant $k = 1/r > 0$. Let $\Sigma \subset \RR^2$ denote the
closed $R$-tube about $C$ of width $R < r$ (see Figure \ref{ex6}). 
Then Theorem \ref{theo1} cannot be applied to a pair of points $z_1, z_2$ 
on $C$.

\medskip

To treat the last two examples we introduce a finer criterion
which is specially adapted to the cases $M = \RR^n$ and $M =
\HH^n$. To do so, we first introduce the {\em spherical area
function} $A: [0,\infty) \to [0,\infty)$ of a triple
$(x,\Omega,M)$ as
\[
A(r) := \vol_{n-1}(S_r(x) \cap \Omega),
\]
where $S_r(x)$ denotes the sphere of radius $r$ about $x$.

\begin{theorem} \label{theo2}
Let $M = \RR^n$ or $M = \HH^n$ and $\Omega_1, \Omega_2 \subset M$ be
two admissible subsets. Let $x_1, x_2 \in M$ and
$A_j: [0,\infty) \to [0,\infty)$ be the corresponding spherical area
functions. If there exist $0 < R < \tilde R$ such that the following
inequalities are satisfied:
\begin{eqnarray*}
A_1(r) &\le& A_2(r) \quad \text{for all $r \in (0,R]$}, \\
A_1(r) &<& A_2(r) \quad \text{for all $r \in (R,\tilde R)$},
\end{eqnarray*}
then $(x_2,\Omega_2,M)$ is initially strictly hotter than
$(x_1,\Omega_1,M)$.
\end{theorem}

Two points $x_1 \in \Omega_1$ and $x_2 \in \Omega_2$ with the same
distance $R > 0$ from the boundaries can thus be compared via the
behavior of the corresponding spherical area functions on the
interval $(R,\tilde R)$. 

\smallskip

\begin{figure}
\begin{center}
\psfrag{u}{$y$} 
\psfrag{t}{$x$} 
\psfrag{t(e)}{$t(\epsilon)$}
\psfrag{c(t)}{$c(t)$}
\psfrag{R+e}{$R+\epsilon$}
\psfrag{th(e)}{$\theta(\epsilon)$}
\psfrag{Sr(P)}{$S_{R+\epsilon}(P)$}
\psfrag{Q}{$Q$}
\psfrag{P=(0,R)}{$P=(0,R)$} 
\includegraphics[height=4cm]{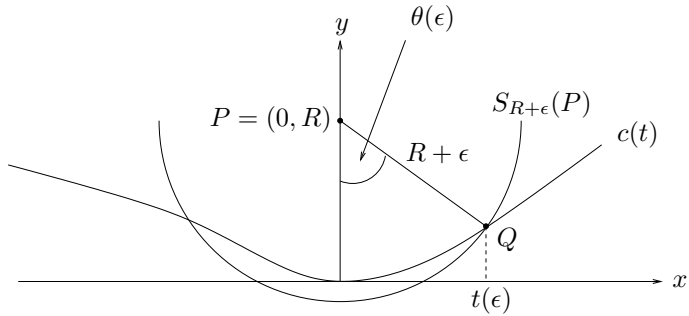}
\end{center}
\caption{Asymptotics of the angle $\theta(\epsilon)$} 
\label{mc}
\end{figure}

Before we return to Examples 5 and 6 we first discuss the asymptotics of
a particular angle (see Figure \ref{mc}) in a useful model case.

\begin{prop} \label{modelas}
Let $\kappa \in \RR$ be a constant and $c: \RR \to \RR^2$ be a 
curve passing horizontally through the origin and given by
\[ c(t) = t (1,0) + \frac{t^2}{2}(0,\kappa+\varphi(t)) \]
with $\lim_{t \to 0 }\varphi(t)=0$. (Note that $\kappa$ is the
curvature of $c$ at $t=0$.) Let $O$ denote the origin and $P$ denote the
point $(0,R) \in \RR^2$ for a fixed $R \in (0,\frac{1}{|k|})$. Then the angle
$\theta(\epsilon) = \angle OPQ$, given by the intersection point $Q$
of the circle $S_{R+\epsilon}(P)$ with the curve $c(\RR)$ near $c(0)$ with
positive horizontal coordinate (see Figure \ref{mc}), has the following
asymptotics,
\[ \theta(\epsilon) = \frac{\sqrt{2 R}}{R+\epsilon}\left( 
\frac{1}{\sqrt{1-R \kappa}} + \psi(\epsilon) \right) \epsilon^{1/2} \]
with $\lim_{\epsilon \to 0^+} \psi(\epsilon) =0$.
\end{prop}

With Theorem \ref{theo2} and Proposition \ref{modelas} in hand, we can
compare the initial heat of the points in Examples 5 and 6.

\medskip

\noindent {\bf Examples 5 and 6 (continued):} In Figure \ref{ex5}
the triple $(x_1,\Omega_1,\RR^2)$ is initially strictly hotter
than $(x_2,\Omega_2,\RR^2)$, since $A_1(r) = A_2(r)$ for $0 < r
\le R$, and $A_1(r) > A_2(r)$ for $R < r < R+\delta$.

In Example 6, let $c: [a,b] \to \RR^2$ be a parametrization of $C$.
We choose a point $z = c(t) \in C$ with corresponding curvature 
$\kappa \in [0,k]$ (the arguments for the case $\kappa \in [-k,0)$ go
analogously). Then $S_R(z) \subset \Sigma$ touches $\partial \Sigma$
in two points where $\partial \Sigma$ has curvatures $0 \le
\frac{\kappa}{1+R \kappa} \le \frac{\kappa}{1-R \kappa}$. 
Proposition \ref{modelas} tells us that the corresponding spherical area 
function satisfies
\[ A(R+\epsilon) = 2\pi(R+\epsilon) - \sqrt{8 R} \left( \sqrt{1-R \kappa} + 
\sqrt{1+R \kappa} + o(1) \right) \epsilon^{1/2}. \] 
(Note that, seen from the point $z$ of the central curve $C$, the proposition 
has to be applied with the curvatures $\kappa_- = -\frac{\kappa}{1-R \kappa}$ 
and $\kappa_+ =\frac{\kappa}{1+R \kappa}$.)

For two points $z_1, z_2 \in C$ with corresponding absolute curvatures
$0 \le \kappa_1 < \kappa_2 \le \kappa$ and corresponding spherical area 
functions $A_i(r)$ we conclude from the concavity of $x \mapsto \sqrt{x}$ that
\[ A_1(R+\epsilon) < A_2(R+\epsilon) \]
for small enough $\epsilon > 0$. Thus, $z_2$ is initially strictly hotter than
$z_1$. The corresponding problem for a tube about a space curve
$C \subset \RR^3$ is discussed in detail in Appendix \ref{spcurve}.

\medskip

The following proposition treats the limiting behavior of the
temperature at boundary points, as $t \to 0$.

\begin{prop} \label{templim}
Let $\Omega \subset M$ be an admissible subset, $x \in \partial \Omega$ 
be a boundary point, $S_r(x)$ be the metric sphere of radius $r > 0$ about 
$x$ and $A: [0,\infty) \to [0,\infty)$ be the associated spherical area
function, i.e., $A(r) = \vol_{n-1}(S_r(x) \cap \Omega)$. 
Assume that the limit on the right hand side of \eqref{splim} exists. Then the
temperature limit at $x$ is given by
\begin{equation} \label{splim}
\lim_{t \to 0^+} f_{\Omega,M}(t,x) = \lim_{r \to 0}
\frac{A(r)}{\vol_{n-1}(S_r(x))}.
\end{equation}
At smooth boundary points $x \in \partial \Omega$, we have, in particular,
\[
\lim_{t \to 0^+} f_{\Omega,M}(t,x) = \frac{1}{2}.
\]
\end{prop}

\noindent {\bf Example 7:} Let $\Pi$ be an arbitrary polygon in the
Euclidean or hyperbolic plane $M$ with angles $\alpha_1, \alpha_2, \dots,
\alpha_n$ at the vertices $x_1, x_2, \dots, x_n$, respectively. Then we
have
\[
\lim_{t \to 0^+} f_{\Delta,M}(t,x) = \begin{cases} 1/2 & \text{if
$x \in \partial \Delta \backslash \{ x_1, x_2, \dots, x_n \}$}, \\
\alpha_j/(2 \pi) & \text{if $x = x_j$}. \end{cases}
\]

\medskip

As a refinement of the boundary behavior at smooth points we have
the following consequence of Theorem \ref{theo2}:

\begin{cor} \label{meancucomp}
Let $\Omega \subset \RR^n$ be an admissible subset and let 
$H_{\partial \Omega}(x)$ denote the {\em mean
curvature} of $\partial \Omega$ at a smooth point $x \in \partial
\Omega$, with respect to the outer unit normal vector. Then we have
\[ H_{\partial \Omega}(z_1) < H_{\partial \Omega}(z_2)
\ \ \Longrightarrow \ \ (z_1,\Omega,\RR^n) \succ
(z_2,\Omega,\RR^n).
\]
\end{cor}

Let us, finally, discuss a heat comparison result based on Steiner
symmetrization in $M = \RR^n$ or $M = \HH^n$. Steiner symmetrization
is a geometric procedure which associates, to every compact set $A
\subset M$, a new set ${\mathcal{S}}(A) \subset M$ of the same volume
which is symmetric with respect to a given hyperplane $E \subset
M$. This geometric procedure has many useful applications in
isoperimetric problems. In this article we use the notions introduced
in \cite{Pey-02}.

\begin{dfn}
Let $M = \RR^n$ or $M = \HH^n$, $g \subset M$ be a geodesic and $E$ be
a orthogonal hyperplane to $g$. $h$ is called a {\em $g$-line} if
there exists a $2$-plane containing both $g$ and $h$ such that $h$ is
a curve of fixed distance to $g$. Let $\pi: M \to E$ denote the
projection whose preimages are the $g$-lines.

Let $A \subset M$ be a compact set. {\em Steiner symmetrization}
${\mathcal{S}}(A) \subset M$ with respect to the data $(g,E)$ is then
uniquely determined by the following properties:
\begin{itemize}
\item[(a)] For all $g$-lines $h$, the intersection ${\mathcal{S}(A)}
\cap h$ is either empty or a bounded closed interval which is
symmetric with respect to $E$.
\item[(b)] If, for every $g$-line $h$, $\lambda_h$ denotes the Riemannian
measure of the submanifold $h \subset M$, then we have
\[ \lambda_h({\mathcal{S}}(A)\cap h) = \lambda_h(A \cap h). \]
Moreover, $ {\mathcal{S}}(A)\cap h$ is empty if and only if $A \cap h$ is
empty.
\end{itemize}
\end{dfn}  

In the Euclidean case, $g$-lines are just straight Euclidean lines
parallel to $g$. If $M = \HH^2$, $g$-lines are hypercycles with the same 
end points as $g$.  

Steiner symmetrization enjoys the following useful properties (see, e.g.,
\cite[Prop. 8]{Pey-02}):

\begin{itemize}
\item[(S1)] $A \subset B$ implies that ${\mathcal{S}}(A) \subset
{\mathcal{S}}(B)$.\\[-3mm]
\item[(S2)] We have $\vol_n({\mathcal{S}}(A)) = \vol_n(A)$. \\[-3mm]
\item[(S3)] If $B \subset M$ is a closed metric ball of radius $R > 0$
about $x \in M$, then ${\mathcal{S}}(B)$ is a closed metric ball
of the same radius about $\pi(x)$. 
\end{itemize}

Our last result reads as follows:

\begin{prop} \label{steinercomp}
Let $M= \RR^n$ or $M = \HH^n$, $\mathcal S$ denote Steiner
symmetrization in $M$ with respect to the data $(g,E)$ and $\pi: M \to
E$ denote orthogonal projection along $g$-lines. Then we have for every
compact admissible set $\Omega \subset M$ and every point
$x \in \Omega$:
\begin{equation} \label{symmcomp}
f_{\Omega,M}(t,x) \le f_{{\mathcal{S}}(\Omega),M}(t,\pi(x)) \quad 
\text{for all times $t > 0$.}
\end{equation}
Let $E_x$ be the orthogonal hyperplane to $g$ through $x$ and $s_x: M
\to M$ denote the reflection in $E_x$. If $\Omega$ is not essentially
symmetric with respect to $E_x$, i.e.,
\[ \vol_n(\Omega\, \Delta\, s_x(\Omega)) > 0, \]
then the above inequality \eqref{symmcomp} holds strictly.
\end{prop}

\begin{figure}[h]
\begin{center}
\psfrag{H2}{{\Large $\HH^2$}} 
\psfrag{g}{$g$} 
\psfrag{E}{$E$}
\psfrag{A}{$A$}
\psfrag{B}{$B$}
\psfrag{C}{$C$}
\psfrag{C0}{$C_0$}
\psfrag{x}{$x$}
\psfrag{p(x)}{$\pi(x)$} 
\includegraphics[width=\textwidth]{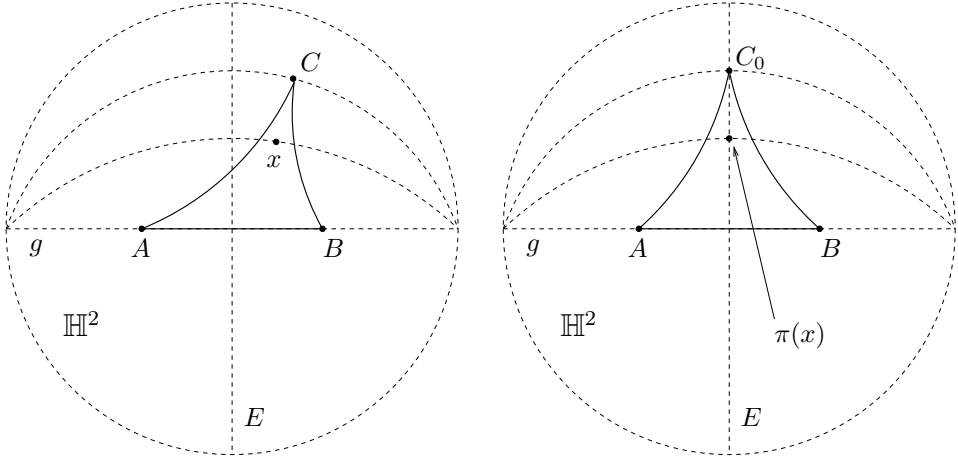}
\end{center}
\caption{Example 8: Heat comparison of hyperbolic triangles with the same 
base and of the same height} 
\label{htss}
\end{figure}

Let us illustrate this result in an example.

\medskip

\noindent {\bf Example 8:} Let $M = \RR^2$ or $M = \HH^2$. Let $AB$ be
a finite interval of a geodesic $g$ in $M$, $E \subset \HH^2$ be the
perpendicular bisector of $AB$ and $\pi: M \to E$ be the
orthogonal projection along $g$-lines. Let $C$ be a point {\em
outside} $E$ and $C_0 = \pi(C) \in E$. The two triangles $\Delta_0 =
\Delta ABC_0$ and $\Delta = \Delta ABC$ have the same base and the
same height and $\Delta_0$ is isosceles (see Figure \ref{htss} for the case
$M = \HH^2$). Then we have for any point $x \in \Delta$:
\begin{equation}\label{strstei}
 f_{\Delta,M}(t,x) < f_{\Delta_0,M}(t,\pi(x)) \quad 
\text{for all times $t > 0$.}
\end{equation}
This can be seen as follows: Proposition \ref{steinercomp} implies that
\[ f_{\Delta,M}(t,x) < f_{{\mathcal{S}}(\Delta),M}(t,\pi(x)) \quad 
\text{for all times $t > 0$.} \]
If $M = \RR^2$, we have ${\mathcal{S}}(\Delta) = \Delta_0$ and
we are done. It remains to consider the case $M = \HH^2$: 
In \cite[Thm. 4]{KP-02}, we proved that ${\mathcal{S}}(\Delta)$ is
strictly contained in $\Delta_0$ (see also \cite{Gue-03} for an
easier proof of this fact). Positivity of the heat kernel implies
strict domain monotonicity of the temperature, i.e., we have
\[ f_{{\mathcal{S}}(\Delta),\HH^2}(t,y) < f_{\Delta_0,\HH^2}(t,y) \quad 
\text{for all $y \in \HH^2$ and all times $t > 0$.} \] 
Choosing $y = \pi(x)$ finishes the proof of inequality \eqref{strstei}
also in the hyperbolic case.

\medskip

A comparison result, based on symmetrization, for solutions of more
general parabolic equations is given, e.g., in \cite{ALT-91}.

\medskip

At the end of the Introduction we like to give a brief explanation of
the structure of this article. In the next section we prove Theorems
\ref{theo1} and \ref{theo2}. Section 3 presents the proofs of all the
other corollaries and propositions of this Introduction. The article ends
with two appendices discussing heat properties in further examples
and an appendix discussing an application of the {\em Principle of not 
feeling the boundary}.

\bigskip

{\sc Acknowledgements:} The authors like to thank Oliver Stein and Djoko
Wiro\-soe\-tisno for helpful discussions.

%%%%%%%%%%%%%%%%%%%%%%%%%%%%%%%%%%%%%%%%%%%%%%%%%%%%%%%%%%%%%%%%%%%

\section{Proof of the main results}

In this section we present the proofs of the two theorems of the
Introduction. In each lemma, proposition and corollary of this section
the geometric requirements on the underlying manifolds are explicitely
stated.

\medskip

All our results are derived from particular properties of the heat
kernel. Classical textbook accounts about heat kernels are, e.g.,
\cite{BGM-72,Cha-84,Gri-99,SchY-94}. Some fundamental properties of
heat kernels are listed in the following proposition.

\begin{prop} \label{hkprops}
Let $M$ be a complete Riemannian manifold with lower
Ricci curvature bound. Then there exists a unique smooth heat kernel
\[ k_M: (0,\infty) \times M \times M \to \RR \] 
with the following properties:
\begin{enumerate}[(HK1)]
\item (positivity) We have $k_M > 0$ on $(0,\infty) \times M \times M$.
\item (heat conservation)
We have, for all $(t,x) \in (0,\infty) \times M$:
\[ \int_M k_M(t,x,y) dy = 1. \]
\item In the case $M = \RR^n$ or $M = \HH^n$ there is a strictly decreasing
function $g_M: [0,\infty) \to \RR$ such that $k_M(t,x,y) = g_M(d(x,y))$.
\item In the case $M = \RR^n$ we have, for any choice $r_1 < r_2
< r_3$ of radii, a constant $\tau_0 > 0$, such that
\[
\int_{M \backslash B_{r_3}(x)} k_M(t,x,y) dy <
\int_{B_{r_2}(x) \backslash B_{r_1}(x)} k_M(t,x,y) dy, \quad
\text{for all $t \in (0,\tau_0)$}.
\]
\end{enumerate}
\end{prop}

\begin{proof}
(HK1), (HK2) and (HK3) are well known facts, see, e.g., 
\cite[pages 181,191,192]{Cha-84}. Multiple applications of integration by
parts yield the inequality
\begin{equation}\label{kr3}
\int_{M \backslash B_{r_3}(x)} k_M(t,x,y) dy \le 
\frac{p(t)}{(4\pi t)^{n/2}} e^{-r_3^2/(4t)}, 
\end{equation}
where $p$ is a polynomial with coefficients only depending on $n$ and $r_3$. 
(If $n$ is odd, \eqref{kr3} holds with equality; if $n$ is even, we first use 
the estimate
\[ \int_{r_3}^\infty r^n e^{-r^2/(4t)} \, dr \le \frac{1}{r_3} 
\int_{r_3}^\infty r^{n+1} e^{-r^2/(4t)}.) \]
On the other hand we have
\begin{equation}\label{kr1r2}
\int_{B_{r_2}(x) \backslash B_{r_1}(x)} k_M(t,x,y) dy \ge 
\frac{C}{(4\pi t)^{n/2}} e^{-r_2^2/(4t)}
\end{equation}
with a fixed constant $C > 0$ only depending on $n,r_1$ and $r_2$. Both
estimates \eqref{kr3} and \eqref{kr1r2} immediately imply property (HK4). 
\end{proof}

\noindent
{\em Remark:} We will show that (HK4) generalizes to arbitrary
Riemannian manifolds, see Corollary \ref{hk4gen} below.  Property
(HK4) is {\em the key observation} in this article.
  
\medskip

\begin{figure}
\begin{center}
\psfrag{x1}{$x_1$} \psfrag{x2}{$x_2$} \psfrag{R}{$R$}
\psfrag{O1}{\LARGE $\Omega_1$} \psfrag{O2}{\LARGE $\Omega_2$}
\includegraphics[width=\textwidth]{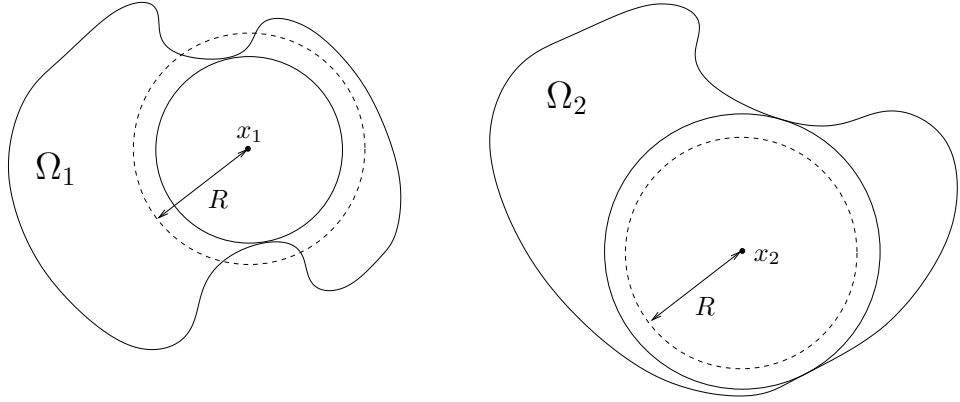}
\end{center}
\caption{Introduction of the comparison triple $(x_1,B_R(x_1),M_1)$}
\label{intro3}
\end{figure}

For the proof of Theorem \ref{theo1} we introduce a third comparison triple
$(x_1,B_R(x_1),M_1)$ and thus break down the statement of the theorem into
two smaller results which are presented in the Propositions A and B below
(see Figure \ref{intro3}). We first state these propositions without proof:

\begin{specprop}{A}
Consider the situation in Theorem \ref{theo1} and let $R \in (R_1,R_2)$. Then 
there exists a $\tau_A > 0$ such that
\[ f_{B_R(x_1),M_1}(x_1,t) < f_{B_{R_2}(x_2),M_2}(x_2,t) \quad
\text{for all $x_1 \in I_1$, $x_2 \in I_2$ and $t \in (0,\tau_A)$.} \]
\end{specprop} 

\begin{specprop}{B}
Consider the situation in Theorem \ref{theo1} and let $R \in (R_1,R_2)$. 
Then there exists a $\tau_B > 0$ such that
\[
f_{\Omega_1,M_1}(x_1,t) < f_{B_R(x_1),M_1}(x_1,t) \quad \text{for all 
$x_1 \in I_1$ and $t \in (0,\tau_B)$.} 
\]
\end{specprop}

\begin{proof}[Proof of Theorem \ref{theo1}] It is sufficient to prove
the general statement of the theorem about the sets $I_1$ and $I_2$. 
We choose $R \in (R_1,R_2)$. Let $0 < t < \min\{ \tau_A, \tau_B\}$ and 
$x_1 \in I_1$ and $x_2 \in I_2$ be given. With the Propositions A and B we 
conclude that
\[ f_{\Omega_1,M_1}(x_1,t) < f_{B_R(x_1),M_1}(x_1,t) < 
f_{B_{R_2}(x_2),M_2}(x_2,t) \le f_{\Omega_2,M_2}(x_2,t),
\]
where the last inequality follows from domain monotonicity $B_{R_2}(x_2)
\subset \Omega_2$. This finishes the proof.
\end{proof}

The proofs of Propositions A and B are a consequence of a sequence of
lemmata, which we discuss next.

\begin{lemma} \label{upperkernel}
Let $M_2$ be a complete Riemannian manifold with lower Ricci curvature
bound $-\kappa < 0$, upper sectional curvature bound $K > 0$ and positive
lower bound $i_0$ on the injectivity radius. Then there exists, for every 
small $\epsilon > 0$, a constant $C_u > 0$ and a time 
$\tau_1 > 0$, both only depending on $\dim M, \kappa, K, i_0, R$ and 
$\epsilon$ such that
\begin{equation} \label{intcompball}
\int_{M_2 \backslash B_R(x)} k_{M_2}(t,x,y) dy \le C_u
e^{-(R-\epsilon)^2/(4 t)}, 
\end{equation}
for all $t \in (0,\tau_1)$ and $x \in M_2$. 
\end{lemma}

\begin{proof}
We assume that $\alpha < 1/4$ is a constant close to $1/4$. We will see later
how $\alpha$ has to be chosen. By the heat kernel estimate of Li and Yau   
(see \cite[Cor. 3.1]{LY-86}), we have
\begin{eqnarray*}
k_M(t,x,y) &\le& \frac{C_1 e^{C_2 t}}
{\left( \vol(B_{\sqrt{t}}(x)) \vol(B_{\sqrt{t}}(y)) \right)^{1/2}}
\exp\left( -\frac{\alpha d^2(x,y)}{t} \right)\\
&\le& \frac{C_3}{t^{n/2}} \exp\left(-\frac{\alpha d^2(x,y)}{t} \right),
\end{eqnarray*}
for all $t \in (0,\sqrt{i_0/2}]$ and $x,y \in M_2$. The constants
$C_1, C_2>0$ depend only on $\alpha, \kappa$ and $n = \dim M$, whereas the
existence of $C_3 >0$ follows from Bishop/G\"unther and depends also
on the upper sectional curvature bound $K > 0$. Now, the volume form
$d\vol = \rho(r,\theta) d\theta dr$ in geodesic polar coordinates
about $x \in M_2$ is defined on a star-shaped subset of $T_xM_2$ and
satisfies
\[
\rho(r,\theta) \le C_4 e^{C_5 r}, \quad \text{with $C_5 =
(n-1)\sqrt{-\kappa}$}.
\]
Estimating the integral in \eqref{intcompball} by an integration
in $T_xM_2$, we obtain
\begin{multline*}
\int_{M_2 \backslash B_R(x)} k_{M_2}(t,x,y) dy \le
\frac{C_3}{t^{n/2}} \int_R^\infty \int_{S^{n-1}}
\exp\left(-\frac{\alpha r^2}{t}\right) \rho(r,\theta) d\theta dr \\
\le \frac{C_6}{t^{n/2}} \int_R^\infty \exp\left(-\frac{\alpha r^2}{t}
+ C_5 r\right) dr \\
\le \frac{C_6}{t^{n/2}} \exp\left( \frac{{C_5}^2 t}{4 \alpha} \right)
\int_R^\infty \frac{r-(C_5 t)/(2 \alpha)}{R-(C_5 t)/(2 \alpha)}
\exp\left(-\frac{\alpha}{t} \left( r-\frac{C_5 t}{2 \alpha} \right)^2
\right) dr \\
\le \frac{C_7}{t^{n/2}} \exp\left(-\frac{\alpha}{t} \left( R-\frac{C_5
t}{2 \alpha} \right)^2 \right).
\end{multline*}
 For $\epsilon > 0$ given, we can choose $0 <
\epsilon_0 < \epsilon$ and $\alpha$ close enough to $1/4$, right
at the beginning, such that there is a time $\tau_1 \in (0,\sqrt{i_0/2})$ 
with
\begin{multline*}
\frac{1}{t^{n/2}} \exp\left(-\frac{\alpha}{t} \left( R-\frac{C_5
t}{2 \alpha} \right)^2 \right) \le \frac{1}{t^{n/2}} \exp\left(
-\frac{(R-\epsilon_0)^2}{4 t} \right) \\
\le \exp\left( -\frac{(R-\epsilon)^2}{4 t} \right), \quad
\text{for all $t \in (0,\tau_1)$.}
\end{multline*}
Note that all constants $C_j > 0$ in this proof are positive and depend
only on the parameters mentioned in the lemma. 
\end{proof}

\begin{lemma} \label{lowerkernel}
Let $M$ be a complete Riemannian manifold with Ricci curvature
bounded from below and $I \subset M$ be a compact subset. Let $R,
\delta >0$ be given such that $R + \delta < \inj(x)$ for all $x \in
I$. Then there exists a constant $C_l > 0$ and a time $\tau_2 > 0$
such that
\begin{equation} \label{lowerannulus}
\int_{B_{R+\delta}(x) \backslash B_R(x)} k_M(t,x,y) dy \ge
C_l e^{-(R+\delta)^2/(4 t)},
\end{equation}
for all $t \in (0,\tau_2)$ and $x \in I$.
\end{lemma}

\begin{proof} Let $U \subset M$ denote the open $(R+2\delta)$-tube about 
$I$ and $k_U^D$ denote the corresponding Dirichlet heat kernel.  
By the Minakshisundaram-Pleijel expansion there is a
smooth function $u_0$ such 
that we have for all $x \in I$ and $y \in B_{R+\delta}(x)$,
\[ k_U^D(t,x,y) = ( u_0(x,y) + O(t) ) \frac{1}{(4
\pi t)^{n/2}} \exp \left( -\frac{d^2(x,y)}{4t} \right), \quad \text{as
$t \to 0$}
\]
with a uniform $O(t)$. Here, $u_0(x,y) = \varphi^{-1/2}(x,y)$ 
(see \cite{Cha-84}), where $\varphi$ is a density function satisfying
\[
\int_{B_R(y)} f(x) dx = \int_0^R \int_{S_yM} f(\exp_y(t v))
\varphi(\exp_y(t v),y) d\vol_{S_yM}(v) dt,
\]
for all $f \in L^1(B_R(y))$.  Note
that, by construction, $k_U^D \le k_M$. Therefore, for all $x, y$ as above 
there exists a $\hat \tau > 0$ such that
\[
k_M(t,x,y) \ge \frac{1}{2} u_0(x,y) \frac{1}{(4 \pi t)^{n/2}} \exp
\left( -\frac{d^2(x,y)}{4t} \right),
\]
for all $t \in (0,\hat \tau)$. It follows that   
\begin{multline*}
\int_{B_{R+\delta}(x) \backslash B_R(x)} k_M(t,x,y) dy =
\int_R^{R+\delta} \int_{S_r(x)} k_M(x,y,t) d\vol_{S_r(x)}(y) dr \\
\ge \frac{1}{(4 \pi t)^{n/2}} \int_R^{R+\delta} e^{-r^2/(4t)}
\int_{S_r(x)} \frac{u_0(x,y)}{2} d\vol_{S_r(x)}(y) dr.
\end{multline*}
Since there exists a $C_0 > 0$ such that
\[ \int_{S_r(x)} u_0(x,y) d\vol_{S_r(x)}(y) dr \ge C_0, \]
for all $x \in I$ and $r \in [R,R+\delta]$, we can find a $\tau_2
\in (0,\hat \tau)$ such that
\[
\int_{B_{R+\delta}(x) \backslash B_R(x)} k_M(t,x,y) dy \ge
\frac{\delta C_0}{2} \frac{e^{-(R+\delta)^2/(4t)}}{(4 \pi
t)^{n/2}} \ge C_l e^{-(R+\delta)^2/4t},
\]
for all $t \in (0,\tau_2)$.
\end{proof}

A consequence of the previous lemmata is the following result, generalizing
property (HK4) to arbitrary Riemannian manifolds.

\begin{cor} \label{hk4gen}
Let $M$ be a complete Riemannian manifold with lower Ricci curvature
bound, upper sectional curvature bound and lower positive bound on
the injectivity radius. Let $I \subset M$ be a compact subset and $0 \le
R_0 < \tilde R$ with $\inj(x) > R_0$ for all $x \in I$. Then there 
exists, for any $0 < \delta < \tilde R - R_0$ and every $\eta > 0$, a 
$\tau_0 > 0$ such that we have 
\[
\int_{M \backslash B_{\tilde R}(x)} k_M(t,x,y) dy < \eta
\int_{B_{R_0 + \delta}(x) \backslash B_{R_0}(x)} k_M(t,x,y) dy,
\]
for all $t \in (0,\tau_0)$ and all $x \in I$.
\end{cor}

\begin{proof}
Without loss of generality, we may assume that $R_0 + \delta < \inj(x)$
for all $x \in I$. Now, choose $\epsilon > 0$ such that $R_0+\delta <
\tilde R - \epsilon$. Then we can conclude with the help of Lemmata
\ref{upperkernel} and \ref{lowerkernel} that there is 
a $\tau_0 \in (0,\min\{\tau_1,\tau_2\})$ such that
\begin{multline*}
\int_{M \backslash B_{\tilde R}(x)} k_M(t,x,y) dy \le
C_u e^{-(\tilde R-\epsilon)^2/(4t)}\\
< \eta C_l e^{-(R_0+\delta)^2/(4t)} \le \eta \int_{B_{R_0 +
\delta}(x) \backslash B_{R_0}(x)} k_M(t,x,y) dy,
\end{multline*}
for all $t \in (0,\tau_0)$ and all $x \in I$.
\end{proof}

Next, we prove Proposition A:

\begin{proof}[Proof of Proposition A]
Choose $\delta, \epsilon> 0$ such
that $R_1 + \delta < \min\{\inj (x_1), R_2 - \epsilon\}$ for all $x_1
\in I_1$. We conclude from (HK2) and Lemmata \ref{upperkernel} and 
\ref{lowerkernel} 
that there is a $\tau_A \in (0,\min\{\tau_1,\tau_2\})$ such that we have, 
for all $x_1 \in I_1$, $x_2 \in I_2$ and $t \in (0,\tau_A)$,
\begin{multline*}
\int_{B_{R_1}(x_1)} k_{M_1}(t,x_1,y) dy = 1 -
\int_{M_1 \backslash B_{R_1}(x_1)} k_{M_1}(t,x_1,y)
dy\\
\le 1 - C_l e^{-(R_1+\delta)^2/(4t)} < 1 - C_u e^{-(R_2-\epsilon)^2/(4 t)}\\
\le 1 - \int_{M_2 \backslash B_{R_2}(x_2)} k_{M_2}(t,x_2,y)dy = 
\int_{B_{R_2}(x_2)} k_{M_2}(t,x_2,y) dy.
\end{multline*}
This finishes the proof of Proposition A.
\end{proof}

The proof of Proposition B is based on the following fact:

\begin{lemma}[Rearrangement-Lemma] \label{rearr}
Assume that there are two non-negative
functions $A_1, A_2: [0,\rho] \to \RR$ satisfying the following
properties:
\begin{itemize}
\item[i)] $A_1(r) \le A_2(r)$, for all $r \in [0,\rho]$,
\item[i))] $\exists\ \rho_0 \in [0,\rho)$ with $\int_0^\rho A_1(r) dr
< \int_0^{\rho_0} A_2(r) dr$.
\end{itemize}
Then we have, for every non-increasing function $f: [0,\rho]
\to (0,\infty)$:
\[ \int_0^\rho A_1(r) f(r) dr < \int_0^{\rho_0} A_2(r) f(r) dr. \]
\end{lemma}

\begin{proof} We have
\begin{multline*}
\int_0^\rho A_1(r) f(r)\, dr = \int_0^{\rho_0} A_1(r) f(r) \, dr +
\int_{\rho_0}^\rho A_1(r) f(r) \, dr \\
=  \int_0^{\rho_0} A_2(r) f(r) \, dr + \int_0^{\rho_0} 
\underbrace{(A_1(r)-A_2(r)}_{\le 0} f(r)\, dr + \int_{\rho_0}^\rho
A_1(r) f(r) \, dr\\
\le \int_0^{\rho_0} A_2(r) f(r) \, dr + f(\rho_0) \int_0^{\rho_0} 
A_1(r)-A_2(r)\, dr + f(\rho_0) \int_{\rho_0}^\rho A_1(r) \, dr\\
= \int_0^{\rho_0} A_2(r) f(r)\, dr + f(\rho_0) \left( \int_0^\rho A_1(r) \,
dr - \int_0^{\rho_0} A_2(r) \, dr \right) < \int_0^{\rho_0} A_2(r) f(r) \, dr.
\end{multline*} 
\end{proof}

As a consequence of the Rearrangement-Lemma we have the following
property of the heat kernel:

\begin{lemma} \label{finomannulus}
Let $M$ be a complete Riemannian manifold with lower bound on the Ricci
curvature, upper bound on the sectional curvature and lower positive bound
on the injectivity radius. Let $\Omega \subset M$ be an admissible set and
$R: M \to [0,\infty]$ be the corresponding distance from the complement. 
Let $I \subset M$ be compact and $R_1, \tilde R > 0$ satisfy 
\[ R(x) \le R_1 < \tilde R < \inj(x) \quad \text{for all $x \in I$.} \]  
Then there exists an $R_0 \in (R_1,\tilde R)$ and a time $\tau_3 > 0$ such
that
\[ \int_{\Omega \cap B_{\tilde R}(x)} k_M(t,x,y) dy <
\int_{B_{R_0}(x)} k_M(t,x,y) dy, \] 
for all $t \in (0,\tau_3)$ and all $x \in I$.
\end{lemma}

\begin{proof} Let
\[ u_t(x,y) := (4\pi t)^{n/2} e^{d^2(x,y)/(4t)} k_M(t,x,y). \]
Applying Corollary \ref{smtas} of Appendix \ref{prnotftb} (with
$\delta = \tilde R$) we have $u_t(x,z) \to u_0(x,z)$, uniformly on
$B_{\tilde R}(x)$, as $t \to 0$. Here, $u_0$ is given by the 
Minakshisundaram-Pleijel expansion. Now we introduce the functions
\begin{eqnarray*}
A_1(t,r) &:=& \int_{S_r(x) \cap \Omega} u_t(x,y) d\vol_{S_r(x)}(y), \\
A_2(t,r) &:=& \int_{S_r(x)} u_t(x,y) d\vol_{S_r(x)}(y),
\end{eqnarray*}
for all small $t \ge 0$. Since $\tilde R > R(x)$, we have
\[ \int_0^{\tilde R} A_1(0,r) dr < \int_0^{\tilde R} A_2(0,r). \]
Choosing an $R_0 \in [R(x),\tilde R)$ with
\[ \int_0^{\tilde R} A_1(0,r) dr < \int_0^{R_0} A_2(0,r) dr, \]
there is also a $\tau_3 > 0$ such that
\[ \int_0^{\tilde R} A_1(t,r) dr < \int_0^{R_0} A_2(t,r) dr, \]
for all $t \in [0,\tau_3)$. Now we apply the Rearrangement Lemma with
the function $f_t(r) = \frac{1}{(4\pi t)^{n/2}} e^{-r^2/(4t)}$ and
obtain
\begin{multline*}
\int_{\Omega \cap B_{\tilde R}(x)} k_M(t,x,y) dy =
\int_0^{\tilde R} \int_{S_r(x) \cap \Omega} k_M(t,x,y) d\vol_{S_r(x)}(y)
dr\\
< \int_0^{R_0} \int_{S_r(x)} k_M(t,x,y) d\vol_{S_r(x)}(y) dr =
\int_{B_{R_0}(x)} k_M(t,x,y) dy,
\end{multline*}
for all $t \in (0,\tau_3)$.
\end{proof}

Now we prove Proposition B: 

\begin{proof}[Proof of Proposition B]
We choose $\tilde R \in (R_1,R)$. We obviously have for all $x_1 \in I_1$:
\[ \int_{\Omega_1} k_{M_1}(t,x_1,y) dy \le \int_{\Omega_1 \cap 
B_{\tilde R}(x_1)} k_{M_1}(t,x_1,y) dy + \int_{M_1 \backslash 
B_{\tilde R}(x_1)} k_{M_1}(t,x_1,y) dy. \]
We conclude from Lemma \ref{finomannulus} that there is an $R_0 \in (R_1,
\tilde R)$ and a time $\tau_3 > 0$ such that we have
\[ \int_{\Omega_1 \cap B_{\tilde R}(x_1)} k_{M_1}(t,x_1,y) dy
< \int_{B_{R_0}(x_1)} k_{M_1}(t,x_1,y) dy,
\] 
for all $x_1 \in I_1$ and $t \in (0,\tau_3)$. Choosing $0 < \delta < 
\tilde R - R_0$ we find, with the help of Corollary \ref{hk4gen}, a time 
$\tau_0 > 0$ such that
\[ \int_{M_1 \backslash B_{\tilde R}(x_1)} k_{M_1}(t,x_1,y) dy <
\int_{B_{R_0 + \delta}(x_1) \backslash B_{R_0}(x_1)} k_{M_1}(t,x_1,y) dy, \]
for all $x_1 \in I_1$ and $t \in (0,\tau_0)$. Combining these facts we end 
up with
\[ \int_{\Omega_1} k_{M_1}(t,x_1,y) dy < \int_{B_{R_0 + \delta}(x_1)}
k_{M_1}(t,x_1,y) dy, \]
for all $x_1 \in I_1$ and $t \in (0,\min\{\tau_0,\tau_3\})$.
Since $R_0 + \delta < \tilde R < R$, this proves Proposition B.
\end{proof}

The above lemmata enable us, finally, to present a relatively short
proof of Theorem \ref{theo2}.

\begin{proof}[Proof of Theorem \ref{theo2}] 
We start with an obvious inequality and use (HK3) to obtain
\begin{multline*}
\int_{\Omega_1} k_M(t,x_1,y)\, dy \le \int_{\Omega_1 \cap B_{\tilde R}(x_1)} 
k_M(t,x_1,y)\, dy + \int_{M \backslash B_{\tilde R}(x_1)} k_M(t,x_1,y)\, dy\\
= \int_0^{\tilde R} g_M(r) A_1(r) \, dr + \int_{M \backslash 
B_{\tilde R}(x_1)} k_M(t,x_1,y)\, dy.
\end{multline*}
The assumptions on $A_1, A_2$ and positivity of the heat kernel imply that 
there is an $R_0 \in (R,\tilde R)$ such
that
\[ \int_0^{\tilde R} g_M(r) A_1(r)\, dr < 
\int_0^{R_0} g_M(r) A_2(r)\, dr.
\] 
Choose $\delta > 0$ such that $R_0 + \delta < \tilde R$. Since $A_2(r) > 0$
for all $r \in [R_0,R_0+\delta]$ we can find an $\eta > 0$ such that
\[ 
A_2(r) = \vol_{n-1}(S_r(x_2)\cap \Omega_2) \ge \eta \vol_{n-1}(S_r(x_2)), 
\quad \text{for all $r \in [R_0,R_0+\delta]$.}
 \]
Corollary \ref{hk4gen} implies that there exists a $\tau_0 > 0$ such that
\begin{multline*}
\int_{M \backslash B_{\tilde R}(x_1)} k_M(t,x_1,y)\, dy < \eta
\int_{R_0}^{R_0+\delta} g_M(r) \vol_{n-1}(S_r(x_2))\, dr\\
\le \int_{R_0}^{R_0+\delta} g_M(r) A_2(r)\, dr,
\end{multline*}
for all $t \in (0,\tau_0)$. Putting these inequalities together we 
conclude that
\[ 
\int_{\Omega_1} k_M(t,x_1,y)\, dy < \int_0^{R_0+\delta} g_M(r) A_2(r)\,
dr \le \int_{\Omega_2} k_M(t,x_2,y)\, dy, \]
for all $t \in (0,\tau_0)$. 
\end{proof}

%%%%%%%%%%%%%%%%%%%%%%%%%%%%%%%%%%%%%%%%%%%%%%%%%%%%%%%%%%%%%%%%%%%%%%

\section{Proof of the other results of the Introduction}

\begin{proof}[Proof of Corollary \ref{theo1comp}]

Let $\Gamma_j = \overline{\Omega_j^c}$ for $j=1,2$. Then $R_j^-$ agree with 
the distances from the complements $\Gamma_j^c$. Applying Theorem
\ref{theo1} we conclude that
\[ (x_1,\Gamma_1,M_1) \succ (x_2,\Gamma_2,M_2). \]
Property (HK2) and $\vol_n(\partial \Omega_j) = 0$ imply
\[ (x_2,\Omega_2,M_2) \succ (x_1,\Omega_1,M_1), \]
finishing the proof.

\end{proof}

Corollary \ref{uniqhot} is a trivial consequence of Theorem \ref{theo1}.

\begin{proof}[Proof of Corollary \ref{conv0}]

Let $f: (0,\infty) \times M \to \RR$ denote the unique solution of 
\eqref{heateq}. 

It is sufficient to prove that the limit of every convergent
subsequence of $x_j$ lies in $I_\infty(\Omega)$. So let us choose a
convergent subsequence which we denote for simplicity, again, by
$x_j$. By continuity of $R: \Omega \to [0,\infty)$ it suffices to
prove that $R(x_j) \to R_\infty(\Omega)$. Let $\epsilon > 0$ be
an arbitrary small number. Choosing $I_1 := \{ x \in \Omega
\mid R(x) \le R_\infty(\Omega) - \epsilon \}$ and $I_2 :=
I_\infty(\Omega)$, we conclude from Theorem \ref{theo1} that there is
a $\tau > 0$ such that we have for all $0 < t < \tau$,
\[ f(t,z_1) < f(t,z_2) \quad \text{for all $z_1 \in I_1$, $z_2 \in I_2$.} \]
Consequently, we have $x_j \not\in I_1$ for all $j$ with $t_j <
\tau$. This finishes the proof of the corollary.
\end{proof}

\begin{proof}[Proof of Proposition \ref{uniqpt}]

The Euclidean heat kernel on $\RR^n$ is given by
\[ k(t,x,y) = \frac{1}{(4\pi t)^{n/2}} e^{-(x-y)^2/(4t)}. \]

Let $\Omega \subset \RR^n$ be compact with $\vol_n(\Omega) > 0$ and 
$\tilde \Omega$ be its convex hull. By \cite{ChK-90} the maxima of
$f := f_{\Omega,\RR^n}$ are located in $\tilde \Omega$. 

We first prove that there is a $T > 0$ such that 
\[ f(t,x) = \int_\Omega k(t,x,y) dy \]
is a concave function on $\tilde \Omega$ for all $t \ge T$. 
This immediately implies that $H(t) \subset \tilde \Omega$ consists of a 
single point for all $t \ge T$. An easy calculation shows that
\[ D^2_x\, k(t,x,y) = \frac{k(t,x,y)}{2t} Q(t,x-y) \quad \text{with}\ 
Q(t,z) = \left( \frac{z_i z_j}{2t} - \delta_{ij} \right)_{ij}. \] 
Note that $Q(t,z)$ can be considered as a perturbation of the negative
definite matrix $- {\rm Id}$. Since $\tilde \Omega$ is bounded, there
is a $T > 0$ such that $Q(t,x-y)$ is negative definite for all $x,y
\in \tilde \Omega$ and all $t \ge T$.  This implies for $t \ge
T$ and $x \in \tilde \Omega$ that
\[ \left\langle D^2_x f(t,x)v , v \right\rangle = \int_\Omega 
\frac{k(t,x,y)}{2t} \underbrace{\left\langle Q(t,x-y) v,v \right\rangle}_{< 0} 
dy < 0, \]
i.e., $D^2_x f(t,x)$ is negative definite.

Since $D_xf: \RR \times \RR^n \to \RR^n$ is smooth, $D_xf(t,H(t)) = 0$ and
$\det D^2_xf(t,H(t)) \neq 0$ for $t \ge T$, the implicit function theorem 
tells us that the map $t \mapsto H(t)$ is smooth on $(T,\infty)$. Finally, 
we have
\begin{multline*}
\frac{\partial}{\partial t} f(t,H(t)) = \frac{\partial f}{\partial t}(t,H(t))
+ \underbrace{\langle D_x f(t,H(t)), \dot H(t) \rangle}_{=0}\\
= \Delta f(t,H(t)) = {\rm tr}\, D^2_xf (t,H(t)) < 0.
\end{multline*} 

\end{proof}

Let $M$ be the Euclidean or the hyperbolic plane. Note that for every $t > 0$ 
we can express the heat kernel $k_M$ by
\[ k_M(t,x,y) = g_M(d(x,y)) \]
with a strictly decreasing function $g_M: [0,\infty) \to \RR$ (see property
(HK3)). Proposition \ref{applmom} follows now immediately from the following
more general result (for a proof see, e.g., \cite{FeT-73} or \cite{Flo-93}):

\begin{theorem}[Momentum lemma]
Let $M$ be the Euclidean or the hyperbolic plane and $g: [0,\infty) \to \RR$ be
a strictly descreasing function. Let $\Sigma_1, \Sigma_2 \subset M$ be two 
polygons of the same area and the same number of sides. Assume that $\Sigma_1$
is regular and that $x_1$ is its center. Then we have for all $x_2 \in M$:
\[ \int_{\Sigma_1} g(d(x_1,x)) dx \ge \int_{\Sigma_2} g(d(x_2,x))
dx, \]
and equality only holds if $\Sigma_2$ is also regular and if $x_2$ is the
center of $\Sigma_2$.
\end{theorem}

\begin{proof}[Proof of Proposition \ref{modelas}]

The situation of the proposition is illustrated in Figure
\ref{mc}. Note that we have $| P - c(t) | > R$ for $t$ close enough to
the origin $0$. Thus the curve $c(\RR)$ and the circle $S_{R+\epsilon}(P)$
intersect in two points, for small enough $\epsilon > 0$, one of which is
denoted by $Q$.

Pythagoras and the asymptotics of $x \mapsto \sqrt{1+x}$ imply that
\[ \epsilon(t) = | P - c(t) | - R = \frac{1-R\kappa}{2R} t^2 + o(t^2). \]
Thus we can also express $t$ as function of $\epsilon$ near $0$ and
obtain
\[ t(\epsilon) = \frac{\sqrt{2 R}}{\sqrt{1-R\kappa}} \epsilon^{1/2} + 
o(\epsilon^{1/2}). \]
This implies that
\[ \theta(\epsilon) \sim \sin \theta(\epsilon) = 
\frac{t(\epsilon)}{R+\epsilon} = \frac{1}{R+\epsilon} \left( 
\frac{\sqrt{2 R}}{\sqrt{1-R\kappa}} \epsilon^{1/2} + 
o(\epsilon^{1/2}) \right), \]
which proves the proposition.
\end{proof}

\begin{proof}[Proof of Proposition \ref{templim}]
Let $n = \dim M$ and $r_0 := \inj(x) > 0$. We conclude from Lemma
\ref{upperkernel} that
\[ \lim_{t \to 0^+} \int_\Omega k_M(t,x,y) dy = 
\lim_{t \to 0^+} \int_{\Omega \cap B_{r_0/2}(x)} k_M(t,x,y) dy. \]
Let $\omega_{n-1}$ denote the volume of the unit sphere in $\RR^n$.
Applying Corollary \ref{smtas} in Appendix \ref{prnotftb} (with $I = \{ x \}$ 
and  $\delta = r_0/2$) we obtain 
\begin{multline*}
\lim_{t \to 0^+} \int_{\Omega \cap B_{r_0/2}(x)} k_M(t,x,y) dy
\\ = \lim_{t \to 0^+} \frac{1}{(4\pi t)^{n/2}} \int_0^{r_0/2} 
e^{-\frac{r^2}{4t}} \int_{S_r(x) \cap \Omega} \left(u_0(x,y) + O(t)
\right)\, d\vol_{S_r(x)}(y) dr,
\end{multline*}
with a uniform $O(t)$ on the compact set $B_{r_0/2}(x) \subset M$. Introducing
\[ g(r) := \frac{\int_{S_r(x) \cap \Omega} u_0(x,y) d\vol_{S_r(x)}(y)}
{\omega_{n-1} r^{n-1}}, \]
we conclude that
\begin{multline*}
\lim_{t \to 0^+} \frac{1}{(4\pi t)^{n/2}} \int_0^{r_0/2} e^{-\frac{r^2}{4t}}
\int_{S_r(x) \cap \Omega} \left(u_0(x,y) + O(t)\right)\, 
d\vol_{S_r(x)}(y) dr \\
= \lim_{t \to 0^+} \frac{1}{(4\pi t)^{n/2}} \int_0^\infty
e^{-\frac{r^2}{4t}} \omega_{n-1} r^{n-1} g(r) dr \\
= \lim_{t \to 0^+} \int_{\RR^n} k_{\RR^n}(t,x,y) g(|x-y|) dy = \lim_{r \to 0}
g(r).
\end{multline*}
The proposition follows now from 
\[
\lim_{r \to 0} \frac{\vol_{n-1}(S_r(x))}{\omega_{n-1} r^{n-1}} = 1
\]
and the fact that $\lim_{r \to 0} u_0(x,\exp_x(r\xi)) = 1$, uniformly for all
$\xi$ in the tangent space $S_xM$.
\end{proof}

\begin{proof}[Proof of Corollary \ref{meancucomp}] The proof proceeds in
two steps.

\noindent {\em Step 1.} We first consider the following model situation: 
Let $C \subset \RR^2$ be a smooth planar curve through
the origin $O$ with horizontal tangent and curvature $k$ at $O$. Then
there exists, locally near $O$, a parametrization $c(r) = (x(r),y(r))$ of $C$ 
with $\dot x(r) > 0$ and satisfying 
\[ |c(r)| = r. \]
This forces the Taylor expansions of the components $x(r)$, $y(r)$ to be
of the form
\begin{eqnarray*}
x(r) &=& r(1+O(r)), \\
y(r) &=& r^2(\frac{k}{2} + O(r)).
\end{eqnarray*}
Let $\theta(r) \in [0,\pi/2)$ denote the angle between the line
through $O$ and $c(t)$ and the horizontal $x$-axis. The above expansions imply
for the asymptotics of the angle $\theta(r)$ that
\[ \theta(r) \sim \tan \theta(r) = \frac{y(r)}{x(r)} = r \left( \frac{k}{2} 
+ O(r) \right). \] 

\begin{figure}
\begin{center}
\psfrag{xi}{$\xi$} 
\psfrag{z}{$z$} 
\psfrag{Sr(z)nO}{$S_r(z) \cap \Omega$}
\psfrag{dO}{$\partial \Omega$}
\psfrag{TzdO}{$T_z\partial \Omega$}
\psfrag{r(pi2-thxi(r))}{$r\left( \frac{\pi}{2}-\theta_\xi(r) \right)$} 
\psfrag{Rv}{$\RR \nu$}
\includegraphics[width=\textwidth]{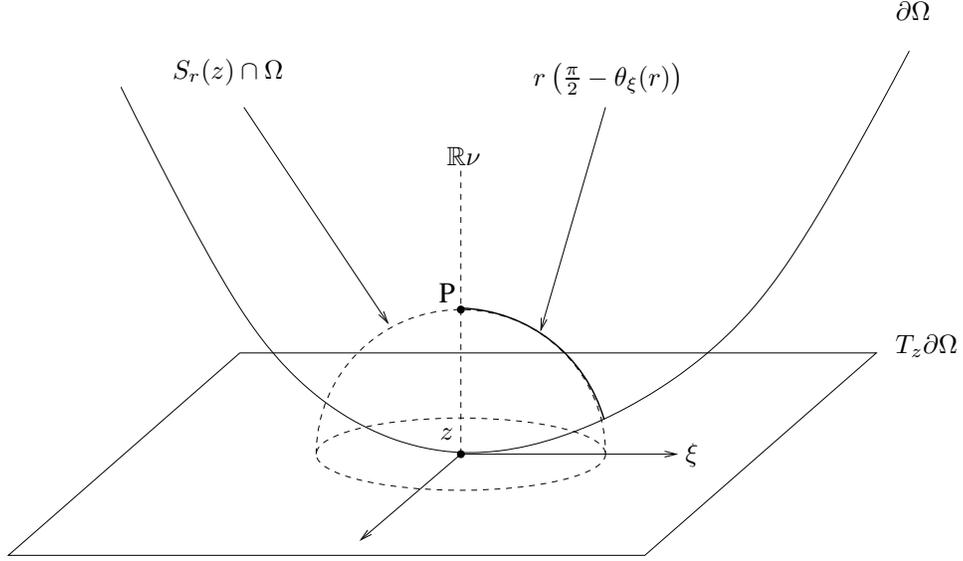}
\end{center}
\caption{Calculation of $\vol_{n-1}(S_r(z) \cap \Omega)$} 
\label{means}
\end{figure}

\noindent {\em Step 2.} Let $z \in \partial \Omega$ be a smooth
boundary point and $L_z: T_z\partial \Omega \to T_z\partial \Omega$
denote the Weingarten map of $\partial \Omega$ at $z$ with respect to
the outward unit normal vector $\nu$ of $\Omega$ in $z$. Using polar
coordinates in $S_r(z)$ about the center $P = z - r \nu \in S_r(z)$
(see Figure \ref{means}) we conclude that
\[ \vol_{n-1}(S_r(z) \cap \Omega) = r^{n-1} \int_{S_z\partial \Omega}
d\xi \int_0^{\frac{\pi}{2}-\theta_\xi(r)} \sin^{n-2}(t)\, dt, \]
where $S_z\partial \Omega$ is the unit tangent space of 
$\partial \Omega$ in $z$ and canonically isometric to the standard
unit sphere $S^{n-2}$ and $\theta_\xi$ is asymptotically given by
\[ \theta_\xi(r) = r \left( \frac{k(\xi)}{2} + O(r) \right), \]
where $k(\xi) = \langle L_z \xi, \xi \rangle$ is the normal curvature
of $\partial \Omega$ at $z$ in direction $\xi$, by Step 1. Since
\[ \int_0^{\frac{\pi}{2}-\theta} \sin^{n-2}(t)\, dt = \begin{cases}
1 - \sin(\theta) + \sum_{j=1}^k a_j (1- \sin^{2j+1}(\theta)) & \text{if 
$n-2 = 2k+1$,}\\[.2cm] 
b (\frac{\pi}{2} - \theta) - \sum_{j=1}^k b_j \sin(2j \theta) 
& \text{if $n-2 = 2k$,} \end{cases}
\]
with suitable constants $a_j, b, b_j$ and $b,b_j > 0$, we conclude that 
\[ \int_0^{\frac{\pi}{2}-\theta_\xi(r)} \sin^{n-2}(t)\, dt = 
C_1 - C_2 r \frac{k(\xi)}{2} + O(r^2), \]
with suitable constants $C_1, C_2 > 0$. This implies that 
\begin{eqnarray*}
A(r) &=& \vol_{n-1}(S_r(z) \cap \Omega) = \frac{1}{2} \vol_{n-1}(S_r(z)) - 
C_2 \frac{r^n}{2} \int_{S_z\partial \Omega} \langle L_z \xi, \xi \rangle d\xi 
+ O(r^{n+1})\\
&=& \frac{\vol_{n-1}(S^{n-1})}{2} r^{n-1} - 
C_2 \frac{\vol_{n-2}(S^{n-2})}{2} H_{\partial \Omega}(z) r^n+ O(r^{n+1}),
\end{eqnarray*}
using $H_{\partial \Omega}(z) = {\rm tr}(L_z)/(n-1)$. This finishes the proof
of the proposition.

\end{proof}

\begin{proof}[Proof of Proposition \ref{steinercomp}] 
Introducing
\[ B_1(r) := \vol_n(B_r(x) \cap \Omega) \quad \text{and} \ B_2(r) = 
\vol_n(B_r(\pi(x)) \cap {\mathcal{S}}(\Omega)),\] 
we first note that
\begin{equation} \label{b1leb2}
B_1(r) \le B_2(r) \quad \text{for all $r \ge 0$.}
\end{equation}
Namely, properties (S1) and (S3) imply that
\[ {\mathcal{S}}(B_r(x) \cap \Omega) \subset B_r(\pi(x)) \cap 
{\mathcal{S}}(\Omega)\, \]
and we obtain \eqref{b1leb2} immediately with the help of property (S2). 
Moreover, we have $B_j(0) =0$ and there is a constant $r_0 > 0$ such that
\[ B_1(r_0) = B_2(r_0) = \vol_n(\Omega). \]
Let $g_M: [0,\infty) \to (0,\infty)$ be the strictly decreasing function
satisfying $k_M(x,y) = g_M(d(x,y))$ for all $x,y \in M$ (see property (HK3)). 
Inequality \eqref{symmcomp} follows now from the following 
integration by parts argument for Stiltjes integrals:
\begin{eqnarray*}
\int_\Omega k_M(x,y)\, dy &=& \int_0^{r_0} g_M(r)\, d B_1(r)\\
&=& g(r_0) B_1(r_0) + \int_0^{r_0} B_1(r)\, d (-g_M(r))\\
&\le& g(r_0) B_2(r_0) + \int_0^{r_0} B_2(r)\, d (-g_M(r))\\
&=& \int_0^{r_0} g_M(r)\, d B_2(r) = \int_{{\mathcal{S}}(\Omega)} k_M(\pi(x),y)
\, dy.
\end{eqnarray*}  
The proof of strict inequality in the non-symmetric case needs some harder
work. W.l.o.g, we can assume that $x = \pi(x)$ and $E_x = E$. 
From the above arguments it suffices to prove that there is a 
non-empty open interval $I$ such that $B_1(r) < B_2(r)$ for all
$r \in I$. $E$ bounds two closed half planes $H_1, H_2 \subset M$. 
For $z \in E$, let $h_z$ denote the $g$-line through $z$ and $\lambda_z$
denote the Riemannian measure of the submanifold $h_z \subset M$. 
We introduce the functions $f_j: E \times [0,\infty) \to [0,\infty)$ by
\[ f_j(z,r) := \lambda_z ( B_r(x) \cap \Omega \cap H_j \cap h_z). \] 
Note that the sets $\Omega_j := \Omega \cap H_j$ can be reconstructed from the 
functions $f_j$ up to measure zero. Thus (essential) non-symmetry of $\Omega$
means that $f_1$ and $f_2$ are different measurable functions. In particular,
there exists a radius $\rho > 0$ such that $f_1(\cdot,\rho), f_2(\cdot,\rho) 
\in L^1(E)$ do not coincide. Let $\lambda$ denote the Riemannian measure of the
hyperplane $E \subset M$. Assume that $f_1(\cdot,\rho) > f_2(\cdot,\rho)$ on 
a set of positive measure. For all $L > 0$, $0 < m < n$ we introduce
the sets
\begin{multline*}
A_{L,m,n} := \big\{ w \in E_x \mid L \le f_1(w,\rho) + f_2(w,\rho) \le L + 
1/n\\
\text{and}\ f_1(w,\rho) \ge f_2(w,\rho) + 1/m \big\}.
\end{multline*}
Our assumption implies that we can find a point $z \in E$ and 
$L > 0$, $0 < m < n$ appropriately such that we have, for every open 
neighborhood $U \subset E$ of $z$:
\[ \lambda( U \cap A_{L,m,n} ) > 0. \]
For any small number $\epsilon >0$ (to be specified later), we can choose
an open interval $I$ and a neighborhood $U$ of $z$ such that we have, for all 
$r \in I$ and $w \in U$:
\[ L - \epsilon < \lambda_w( B_r(x) \cap h_w ) < L. \]
Then we have, for all $w \in U \cap A_{L,m,n}$ and all $r \in I$,
\[
\lambda_w \Big( B_r(x) \cap {\mathcal{S}}(\Omega) \cap h_w \Big) = 
\lambda_w( B_r(x) \cap h_w ).
\]
Since $f_2(w,\rho) \le \frac{L}{2} + \frac{1}{2n} - \frac{1}{2m}$ we obtain, 
on the other hand,
\begin{eqnarray*}
\lambda_w \Big( B_r(x) \cap \Omega \cap h_w \Big) &=& f_1(w,r) + f_2(w,r)
\ \le \ \frac{1}{2} \lambda_w( B_r(x) \cap h_w ) + f_2(w,\rho)\\
&<& \lambda_w( B_r(x) \cap h_w ) + \Big( \frac{1}{2n} + \frac{1}{2}\epsilon
\big) - \frac{1}{2m}.
\end{eqnarray*}
Since $m < n$ we can choose $\epsilon > 0$ originally small enough such that
\[ \lambda_w \Big( B_r(x) \cap \Omega \cap h_w \Big) < \lambda_w 
 \Big( B_r(x) \cap {\mathcal{S}}(\Omega) \cap h_w \Big) - \delta, \]
for a suitable small $\delta > 0$, for all $r \in I$, on a set of positive 
measure in $U$. Since we have, for all other $w \in E$:
\[ \lambda_w \Big( B_r(x) \cap \Omega \cap h_w \Big) \le \lambda_w 
 \Big( B_r(x) \cap {\mathcal{S}}(\Omega) \cap h_w \Big), \] 
we conclude with Fubini that $B_1(r) < B_2(r)$ for all $r \in I$. This finishes
the proof of strict inequality in the non-symmetric case.
\end{proof}

%%%%%%%%%%%%%%%%%%%%%%%%%%%%%%%%%%%%%%%%%%%%%%%%%%%%%%%%%%%%%%%%%%%%

\appendix

\section{A square-shaped dumbbell}
\label{sqdumbbell}

\begin{figure}
\begin{center}
\psfrag{a}{$a$} 
\psfrag{-a}{$-a$} 
\psfrag{b}{$b$}
\psfrag{-b}{$-b$}
\psfrag{c}{$c$}
\psfrag{-c}{$-c$} 
\psfrag{x}{$x$}
\psfrag{y}{$y$}
\psfrag{b+2a}{$b+2a$}
\psfrag{-b-2a}{$-b-2a$}
\psfrag{O}{$\Omega$}
\psfrag{-x(t)}{$-x(t)$}
\psfrag{x(t)}{$x(t)$}
\includegraphics[height=6cm]{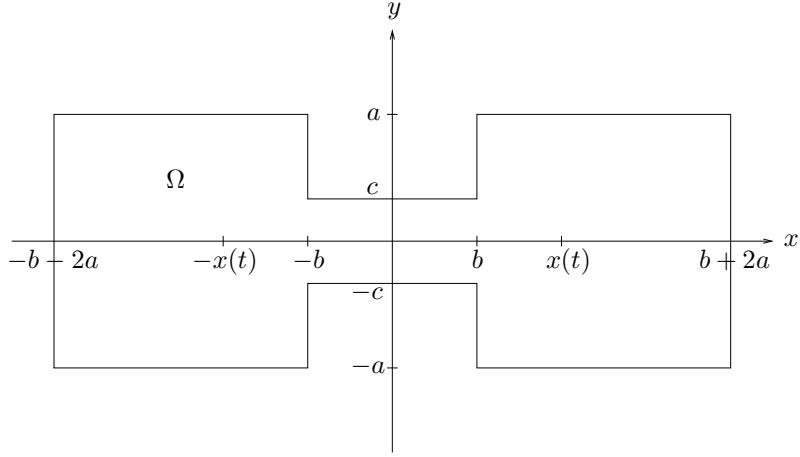}
\end{center}
\caption{Movement of hottest points in $\Omega$} 
\label{sqdb}
\end{figure}

Consider the initial temperature distribution $\chi_\Omega$ where
$\Omega$ consists of two squares of the same side length $a > 0$
connected by a rectangle with the side lengths $b,c > 0$ (see Figure
\ref{sqdb}). One easily concludes that the set $H(t)$ is contained in the
horizontal $x$-axis. The temperature distribution $f(t,x)$
of a point $z = (x,0)$ at time $t > 0$ is given by
\begin{eqnarray*}
2 \pi t f(t,x) &=& \int_0^c e^{-y^2/(4t)} dy
\int_0^{b+2a} e^{-(x-x')^2/(4t)} + e^{-(x+x')^2/(4t)} dx'\\
&+& \int_c^a e^{-y^2/(4t)} dy \int_b^{b+2a} e^{-(x-x')^2/(4t)} 
+ e^{-(x+x')^2/(4t)} dx'.
\end{eqnarray*}
For fixed $t > 0$, we have
\begin{multline*}
\pi\, e^{x^2/(4t)}\, \frac{\partial f}{\partial x}(t,x) =
\left( \int_c^a e^{y^2/(4t)} dy\right) e^{-b^2/(4t)} \sinh(\frac{b}{2t}x) \\
- \left(\int_0^a e^{y^2/(4t)} dy\right) e^{-(b+2a)^2/(4t)} 
\sinh(\frac{(b+2a)}{2t}x).
\end{multline*}
Introducing
\[ h_t(x) := \frac{\sinh(\frac{b}{2t}x)}{\sinh(\frac{b+2a}{2t}x)}, \]
the extrema of $f(t,\cdot)$ are at $x=0$ and at
the $x$-solutions of
\begin{equation} \label{hscond}
h_t(x) = \frac{\int_0^a e^{-y^2/(4t)} dy}{\int_c^a e^{-y^2/(4t)} dy} 
e^{-(a^2+ab)/t}.
\end{equation}
For fixed $t > 0$, the function $h_t$ is even and strictly
decreasing to zero on $[0,\infty)$ with $h_t(0) = b/(b+2a)$.
Consequently, the condition \eqref{hscond} has precisely one
solution $x(t)$ in $(0,\infty)$ iff 
\[ g(t) := \frac{\int_0^a e^{-y^2/(4t)} dy}{\int_c^a e^{-y^2/(4t)} dy} 
e^{-(a^2+ab)/t} < \frac{b}{b+2a}, \]
and no solution, otherwise. Moreover, in the first case we can conclude 
that $x(t) \in (0,b+a)$, because for $x_1= b+a$ we obtain
\[ h_t(x_1) < e^{\frac{b}{2t}x_1 - \frac{b+2a}{2t}x_1} = 
e^{-(a^2+ab)/t} \le g(t), \]
where we used the estimate
\[ \frac{\sinh u}{\sinh v} < e^{u-v} \quad \text{for all $0 \le u < v$}. \] 

Note that
\[ \lim_{t \to 0} g(t) = 0, \quad \lim_{t \to \infty} g(t) = \frac{a}{a-c} >
\frac{b}{b+2a}, \]
and that $g: [0,\infty) \to [0,\infty)$ is a product
\[ g(t) = \frac{1}{\int_c^a e^{(4a^2-y^2)/(4t)} dy} \cdot \int_0^a 
e^{-y^2/(4t)} dy \cdot e^{-ab/t} \] 
of three strictly increasing continuous functions. Therefore there is
a unique $t_0 > 0$ such that
\begin{equation} \label{t0}
 g(t_0) = \frac{\int_0^a e^{-y^2/(4t_0)} dy}{\int_c^a e^{-y^2/(4t_0)} dy} 
e^{-(a^2+ab)/t_0} = \frac{b}{b+2a}.
\end{equation}
We conclude that the set $H(t)$ of hottest points at time $t$ is given by 
$\{-x(t),x(t) \} \subset (-b-a,b+a)$ for $0 < t < t_0$ and that both 
temperature maxima collapse at the origin at time $t = t_0$. For $t > t_0$, 
the origin is the only temperature maximum. Equation \eqref{t0} allows to 
calculate the critical time $t_0$ up to any precision. In the case $c=0$ we
obtain 
\[ t_0 = \frac{1}{4} \frac{(b+2a)^2-b^2}{\ln(b+2a)-\ln(b)}. \]
For fixed $a,b > 0$ the time of collapse $t_0 = t_0(a,b,c)$ becomes
arbitrarily small as $c \nearrow a$.

%%%%%%%%%%%%%%%%%%%%%%%%%%%%%%%%%%%%%%%%%%%%%%%%%%%%%%%%%%%%%%%%%%%%%

\section{A tube about a space curve}
\label{spcurve}

Let $C \subset \RR^3$ be a smooth space curve, $c: [a,b] \to
\RR^3$ be an arc length parametrization of $C$, 
\begin{eqnarray*}
f_1(t) &:=& \dot c(t), \\
f_2(t) &:=& \frac{\ddot c(t)}{\Vert \ddot c(t) \Vert}, \\
f_3(t) &:=& f_1(t) \times f_2(t)
\end{eqnarray*}
be the accompagnying Frenet trihedron and $\kappa, \tau$ the
curvature and the torsion of $c$. Let $0 < R < \frac{1}{\max \kappa(t)}$
and $\Sigma$ be the closed tube of radius $R$ about $C$. 

We fix $t_0 \in [a,b]$ and set $\kappa_0 := \kappa(t_0)$. For every angle 
$\alpha \in [0,2\pi)$ we introduce the plane
\[ E_\alpha = \RR g_1(\alpha) + \RR g_2(\alpha) \]
 with 
\[ g_1(\alpha) = f_1(t_0) \quad \text{and}\ g_2(\alpha) = \cos \alpha 
f_2(t_0) + \sin \alpha f_3(t_0). \] 
There is a canonical identification of $E_\alpha$ with $\RR^2$ via
$(x,y) \mapsto x g_1(\alpha) + y g_2(\alpha)$. 

\begin{figure}
\begin{center}
\psfrag{Ea}{$c(t_0) + E_\alpha$} 
\psfrag{R}{$R$} 
\psfrag{Ca}{$C_\alpha$}
\psfrag{C}{$C$}
\psfrag{c(t0)}{$c(t_0)$}
\psfrag{z+}{$z_+$} 
\psfrag{g1a}{$g_1(\alpha)$}
\psfrag{g2a}{$g_2(\alpha)$}
\includegraphics[width=\textwidth]{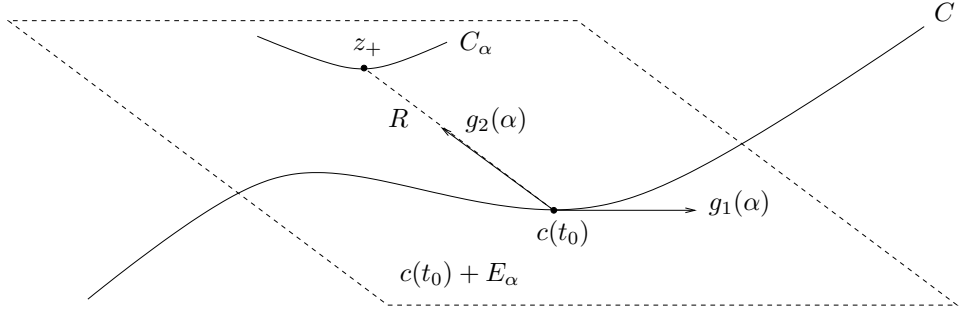}
\end{center}
\caption{The distance $R$ curve $C_\alpha$ of $C$ in direction $\alpha$} 
\label{drc}
\end{figure}

Let $z_\pm := c(t_0) \pm R g_2(t_0) \in \partial \Sigma$. The intersection
$\partial \Sigma \cap (c(t_0) + E_\alpha)$ consists, locally near $z_-$ and 
$z_+$, of two plane curves. Let $C_\alpha \subset c(t_0) + E_\alpha$ denote 
the component of this intersection through $z_+$. We call $C_\alpha$ the 
{\em distance $R$ curve of $C$ in direction $\alpha$} 
(see Figure \ref{drc}). We first prove the following result:

\begin{prop} \label{distabl}
Let $C_\alpha$ be the distance $R$ curve of $C$ in direction
$\alpha \in [0,2\pi)$ and $c_\alpha: (t_0 - \epsilon, t_0 + \epsilon) 
\to \RR^3$ be a parametrization of $C_\alpha$ near $z_+ = c(t_0) + R g_2(t_0)$
satisfying
\begin{equation} \label{Rdist}
 | c(t) - c_\alpha(t) | = R \quad \text{for all $t \in (t_0-\epsilon,
t_0 + \epsilon)$.}
\end{equation}
Let $\kappa_0 \ge 0$ denote the curvature of $C$ in $c(t_0)$. Then we have
\begin{eqnarray*}
\dot c_\alpha(t_0) &=& \big(1 - R \kappa_0 \cos \alpha\big)\, g_1(\alpha), \\
\ddot c_\alpha(t_0) &=& C_1\, g_1(\alpha) + \kappa_0 \cos \alpha\, \big(1-
R \kappa_0 \cos \alpha\big)\, g_2(\alpha).
\end{eqnarray*}
(The constant $C_1$ is of no importance for our considerations.) 
\end{prop}

\begin{proof} 
Let $\alpha \in [0,2\pi)$ be fixed. Let $x, y: (t_0 - \epsilon, t_0 +
\epsilon) \to \RR$ denote the coordinate functions of the curve
$c_\alpha$ in the plane $c(t_0) + E_\alpha$, i.e.,
\begin{eqnarray} \nonumber
c_\alpha(t) &=& c(t_0) + x(t) g_1(\alpha) + y(t) g_2(\alpha)\\
&=& c(t_0) + x(t) f_1(t_0) + y(t)\, \big(\cos \alpha f_2(t_0) + \sin \alpha 
f_3(t_0)\big). \label{ca1}
\end{eqnarray}
Note that $x(0) = 0$ and $y(0) = R$.

\eqref{Rdist} implies that there is a smooth function $\alpha: (t_0 -
\epsilon, t_0 + \epsilon) \to \RR$ with $\alpha(0) = \alpha$ and
\begin{equation} \label{ca2}
c_\alpha(t) = c(t) + R\, \big(\cos \alpha(t)\, f_2(t) + \sin \alpha(t)\, 
f_3(t)\big).
\end{equation}
Using \eqref{ca1}, \eqref{ca2} and the Frenet equations, we conclude that
\begin{multline*}
\dot c_\alpha(t) = \dot x(t) f_1(t_0)\ +\ \dot y(t)\, \big( \cos \alpha 
f_2(t_0) + \sin \alpha f_3(t_0)\big) \\
= \big(1- R \kappa(t) \cos \alpha(t)\big) f_1(t)\ +\ R \,
(\dot \alpha(t) + \tau(t))\, \big( - \sin \alpha(t) f_2(t) + \cos \alpha(t) 
f_3(t) \big).
\end{multline*}
Comparison of coefficients at $t=t_0$ implies that
\[ \dot \alpha(t_0) + \tau(t_0) = 0, \] 
i.e., we have
\[ \dot c_\alpha(t_0) = \big(1-R \kappa_0 \cos \alpha\big)\, g_1(\alpha). \]  
Differentiating again and using again the Frenet equations, we obtain
\begin{eqnarray}
\ddot c_\alpha(t) &=& \ddot x(t) f_1(t_0)\ +\ \ddot y(t)\, \big( \cos \alpha 
f_2(t_0) + \sin \alpha f_3(t_0)\big) \nonumber \\
&=& \big(-R \dot \kappa(t) \cos \alpha(t) + R \kappa(t) \dot \alpha(t) \sin 
\alpha(t) \big)\, f_1(t) \nonumber\\
&& + \big( \kappa(t) - R \kappa^2(t) \cos \alpha(t) - R (\ddot \alpha(t) + \dot
\tau(t)) \sin \alpha(t) \big) \, f_2(t) \label{c2d} \\
&& + R \big( \ddot \alpha(t) + \dot \tau(t) \big)\cos \alpha(t) \, f_3(t)
+ \big( \dot \alpha(t) + \tau(t) \big) \, v(t) \nonumber,
\end{eqnarray}
with a suitable vector valued function $v:  (t_0 - \epsilon, t_0 + \epsilon) 
\to \RR^3$. Using $\dot \alpha(t_0) + \tau(t_0) =0$, the comparison of the 
coefficients of $f_2$ and $f_3$ at $t=t_0$ yields
\[ R (\ddot \alpha(t_0) + \dot \tau(t_0) ) = \kappa_0 \sin \alpha\, \big(1 - R 
\kappa_0 \cos \alpha\big). \] 
Inserting this back into  \eqref{c2d} we end up with
\begin{multline*}
\ddot c_\alpha(t_0) = C_1\, f_1(t_0)\ + \ (1-R \kappa_0 \cos \alpha) 
(\kappa_0 - \kappa_0 \sin^2 \alpha)\, f_2(t_0) \\
+ (1 - R \kappa_0 \cos \alpha) \kappa_0 \sin\alpha \cos \alpha \, f_3(t_0) \\
= C_1\, g_1(\alpha) + (1 - R \kappa_0 \cos \alpha) \kappa_0 \cos \alpha\, 
g_2(\alpha).   
\end{multline*}
This finishes the proof of the proposition.
\end{proof}

Next we calculate the planar curvature $\hat \kappa_\alpha$ of 
$C_\alpha \subset c(t_0) + E_\alpha$ in $z_+$. Let 
\[ g_3(\alpha) = g_1(\alpha) \times g_2(\alpha) = -\sin \alpha f_2 +
\cos \alpha f_3\ \bot\ E_\alpha \]
and
\[ J: E_\alpha \to E_\alpha, \quad J(v) = g_3(\alpha) \times v \]
be rotation in $E_\alpha$ by $\pi/2$. Then Proposition \ref{distabl} 
implies that the planar curvature of $c_\alpha$ in $t_0$ is given by
\[ \hat \kappa_\alpha = \frac{\langle \ddot c_\alpha(t_0), J(\dot 
c_\alpha(t_0)) \rangle}{\Vert \dot c_\alpha(t_0) \Vert^3} = 
\frac{\kappa_0 \cos \alpha}{1- R \kappa_0 \cos \alpha}. \]

\begin{figure}
\begin{center}
\psfrag{c(t0)+Ea}{$c(t_0) + E_\alpha$} 
\psfrag{R+e}{$R+\epsilon$} 
\psfrag{Rg2(a)}{$R g_2(\alpha)$}
\psfrag{g1(a)}{$g_1(\alpha)$}
\psfrag{Ca}{$C_\alpha$}
\psfrag{c(t0)}{$c(t_0)$}
\psfrag{SR+e(c(t0))}{$S_{R+\epsilon}(c(t_0))$} 
\psfrag{ta(e)}{$\theta_\alpha(\epsilon)$}
\includegraphics[width=\textwidth]{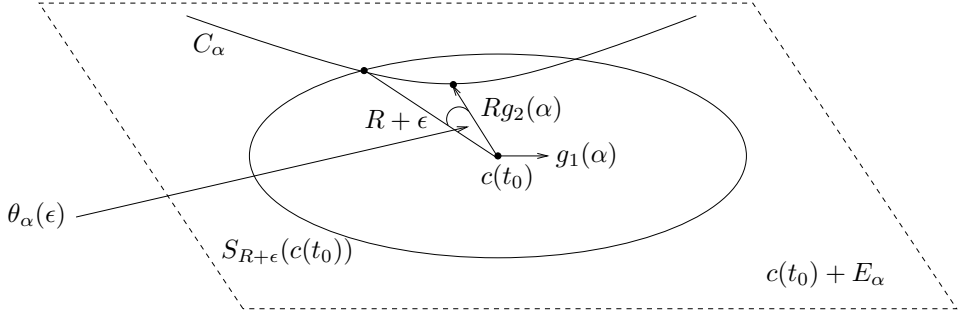}
\end{center}
\caption{The angle $\theta_\alpha(\epsilon)$} 
\label{drc2}
\end{figure}

We are now able to calculate the asymptotics of 
\[ \epsilon \mapsto A(R+\epsilon) = \vol_2(S_{R+\epsilon}(c(t_0)) 
\cap \Sigma). \]  
Using Fermi coordinates on the sphere $S_{R+\epsilon}(c(t_0))$ about the 
great circle
\[ S_{R+\epsilon}(c(t_0))\, \cap\, \left(c(t_0) + \RR g_2(\alpha) + \RR 
g_3(\alpha)\right) \]
we obtain
\begin{eqnarray*}
A(R+\epsilon) &=& 4 \pi (R+\epsilon)^2 - (R+\epsilon)^2 \int_0^{2\pi} 
\int_{-\theta_\alpha(\epsilon)}^{\theta_\alpha(\epsilon)} \cos \theta\, 
d \theta \, d \alpha \\
&=& 4\pi(R+\epsilon)^2 - 2 (R+\epsilon)^2 \int_0^\pi \big( \sin 
\theta_\alpha(\epsilon) + \sin \theta_{\alpha+\pi}(\epsilon) \big) \, 
d\alpha,
\end{eqnarray*} 
where the angle $\theta_\alpha(\epsilon) \in [0,\pi/2]$ is described
in Figure \ref{drc2}. (Note that in Figure \ref{drc2},
$S_{R+\epsilon}(c(t_0))$ denotes the circle about $c(t_0)$ of radius
$R+\epsilon$ in the plane $c(t_0) + E_\alpha$ and not the
$2$-dimensional sphere.) Applying Proposition \ref{modelas} (with
curvature $\kappa = - \hat \kappa_\alpha = - \kappa_0 \cos
\alpha/(1-R\kappa_0 \cos \alpha)$, seen from $c(t_0)$) we conclude
that
\[ \sin \theta_\alpha(\epsilon) \sim \theta_\alpha(\epsilon) = 
\frac{\sqrt{2R}}{R+\epsilon} \left(\sqrt{1-R \kappa_0 \cos \alpha} + o(1) 
\right) \epsilon^{1/2}.
\]
Consequently, we obtain
\begin{multline*}
A(R+\epsilon) = 4\pi(R+\epsilon)^2\\ 
- \sqrt{8R}(R+\epsilon) \left(
\int_0^\pi \sqrt{1-R\kappa_0\cos\alpha} + \sqrt{1+R\kappa_0\cos\alpha} 
\, d\alpha + o(1) \right) \epsilon^{1/2},
\end{multline*}
which implies that 
\[ 0 \le \kappa(t_1) < \kappa(t_2) \ \ \Longrightarrow \ \
(c(t_2),\Sigma,\RR^3) \succ (c(t_1),\Sigma,\RR^3), \] 
similarly as in the two-dimensional case.

%%%%%%%%%%%%%%%%%%%%%%%%%%%%%%%%%%%%%%%%%%%%%%%%%%%%%%%%%%%%%%%%%%%%%

\section{Small time asymptotics of the heat kernel}
\label{prnotftb}

The aim of this appendix is to prove Corollary \ref{smtas} below, a result,
which was used in earlier sections. Let $M$ be a complete Riemannian
manifold with lower bound $- \kappa < 0$ on the Ricci curvature, upper
bound $K > 0$ on the sectional curvature and lower positive bound
$i_0$ on the injectivity radius. For any open subset $U \subset M$ let
$k_U^D(t,x,y)$ denote the corresponding Dirichlet heat kernel.

\begin{theorem}[Principle of not feeling the boundary] \label{pnfb} 
Let $I \subset M$ be a compact subset and $U \subset M$ be an open subset 
containing $I$ and $\epsilon > 0$. Then there exists a constant $C > 0$, 
depending only on $\dim M, \kappa, K$, and $i_0$, such that we have 
for all $x,y \in I$ and all $t \in (0,\sqrt{i_0/2}]$:
\[ | k_M(t,x,y) - k_U^D(t,x,y) | \le C e^{-d^2/(4+\epsilon)t}, \]
where $d \in (0,\infty]$ denotes the distance between $I$ and $\partial U$.
\end{theorem}

\begin{proof}
Let $x,y \in I$ and $t \in (0,\sqrt{i_0/2}]$ be fixed and, for small 
$\delta > 0$, let $f_\delta \in C^\infty(M)$ be a non-negative 
function with support in $B_\delta(y) \subset U$ and total
integral one. Let
\[ g(t,x) = \int_U k_M(t,x,z) f_\delta(z)\, dz \quad \text{and} \
g^D(t,x) = \int_U k_U^D(t,x,z) f_\delta(z)\, dz. \]
Since $k_M \ge k_U^D$, we conclude that 
\[ h = g - g^D: [0,\infty) \times \overline{U} \to [0,\infty) \] 
is a solution of the heat equation with $h(0,x) = 0$ for all $x \in 
\overline{U}$ and $h(s,x) \ge 0$ for all $(s,x) \in (0,\infty) 
\times \partial U$. The maximum principle (see, e.g., 
\cite[Section 6.1]{Tay-96}) implies that
\begin{equation} \label{gtx}
h(t,x) \le \max_{(s,w) \in (0,t] \times \partial U} h(s,w).
\end{equation}
For every $(s,w) \in (0,t] \times \partial U$ we have with 
\cite[Cor.3.1]{LY-86}
\begin{multline*}
0 \le h(s,w) = \int_U \big( k_M(s,w,z) - k_U^D(s,w,z) \big) f_\delta(z)\,
dz \le \int_{B_\delta(y)} k_M(s,w,z) f_\delta(z)\, dz
\\ \le \max_{z \in B_\delta(y)} k_M(s,w,z) \le C \exp\left(\frac{-(d-\delta)^2}
{(4+\epsilon)t}\right), 
\end{multline*} 
since $y \in I, w \in \partial U$ and $d(B_\delta(y),\partial U) \ge d -
\delta$. Note that $C > 0$ depends only on the parameters mentioned
in the theorem.  Letting $\delta \to 0$ we finally derive from
\eqref{gtx} that
\[ 0 \le k_M(t,x,y) - k_U^D(t,x,y) \le C \exp\left(\frac{-d^2}{(4+\epsilon)t}
\right), 
\] 
finishing the proof of the theorem.
\end{proof}

\begin{cor} \label{smtas}
Let $I \subset M$ be a compact subset and $0 < \delta < \min_{x \in I}
\inj(x)$. Then we have, for all $x \in I$ and $y \in B_\delta(x)$:
\[ k_M(t,x,y) = \frac{1}{(4\pi t)^{n/2}} e^{-d^2(x,y)/(4t)} \big( 
u_0(x,y) + O(t) \big),
\]
with a uniform $O(t)$ and $u_0$ given by the Minakshisundaram-Pleijel
expansion.
\end{cor}

\begin{proof}
We choose an open set $U \subset M$ with $d(I,\partial U) > 0$ large enough.
The Minakshisundaram-Pleijel expansion for the Dirichlet heat kernel $k_U^D$
yields, for all $x \in I$ and $y \in B_\delta(x)$:
\[ k_U^D(t,x,y) = \frac{1}{(4\pi t)^{n/2}} e^{-d^2(x,y)/(4t)} \big( 
u_0(x,y) + O(t) \big).
\]
We conclude with Theorem \ref{pnfb} that the same asymptotics holds true for
the heat kernel $k_M$. 
\end{proof}

%%%%%%%%%%%%%%%%%%%%%%%%%%%%%%%%%%%%%%%%%%%%%%%%%%%%%%%%%%%%%%%%%%%%%

\bigskip

{\small

  {\sc Leon Karp}\\ Department of Mathematics, CUNY Graduate Center and
  Lehman College\\ 365 Fifth Avenue, New York, NY 10016, USA\\ 
  email: lkarp@gc.cuny.edu\\

  {\sc Norbert Peyerimhoff}\\ 
  Department of Mathematical Sciences, Science Laboratories\\
  South Road, Durham, DH1 3LE, United Kingdom\\ 
  email: norbert.peyerimhoff@durham.ac.uk

}

\end{document}